\documentclass[fleqn,usenatbib]{mnras}
\usepackage{amssymb}
\usepackage{xcolor} 
\usepackage{booktabs}
\usepackage{array}
\usepackage{pifont}
\usepackage{times}
\usepackage{bm}
\usepackage{amsmath}
\usepackage{latexsym}%
\usepackage{placeins}
\usepackage{graphicx}%
\usepackage{tikz,xcolor,hyperref}
\usetikzlibrary{arrows.meta,decorations.markings,positioning,calc}
\definecolor{lime}{HTML}{A6CE39}
\usepackage{mathrsfs}
\DeclareRobustCommand{\orcidicon}{
\begin{tikzpicture}
\draw[lime, fill=lime] (0,0)
circle[radius=0.16]
node[white]{{\fontfamily{qag}\selectfont \tiny \.{I}D}}; 
\end{tikzpicture}
\hspace{-2mm}
}

\definecolor{tealblue}{rgb}{0.21, 0.46, 0.53}

\usepackage{mathtools}

\usepackage{algorithm}
\usepackage{algpseudocode}

\newcommand{\bbR}{\mathbb{R}}
\newcommand{\cO}{\mathcal{O}}
\newcommand{\frg}{\mathfrak{g}}
\newcommand{\ev}{\mathrm{ev}}

\title[Geometric Fingerprints of Line Profiles]{The Hidden Geometry of Astrophysical Spectra: Path-Signatures of Line Profiles}

\foreach \x in {A, ..., Z}{%
\expandafter\xdef\csname orcid\x\endcsname{\noexpand\href{https://orcid.org/\csname orcidauthor\x\endcsname}{\noexpand\orcidicon}}
}

\author[Rafael S. de Souza \& Severin Bunk]{
Rafael S. de Souza\,\orcidA{}$^{1,2,3,\ddagger}$
\thanks{E-mail: \href{mailto:rd23aag@herts.ac.uk}{rd23aag@herts.ac.uk}}
\quad
Severin Bunk\,\orcidB{}$^{1,\ddagger}$
\thanks{E-mail: \href{mailto:s.bunk@herts.ac.uk}{s.bunk@herts.ac.uk}}
\\
$^{1}$Centre for Astrophysics Research, University of Hertfordshire, College Lane, Hatfield, AL10~9AB, UK\\
$^{2}$Instituto de Física, Universidade Federal do Rio Grande do Sul, Porto Alegre, RS 90040-060, Brazil\\
$^{3}$Department of Physics \& Astronomy, University of North Carolina at Chapel Hill, NC 27599-3255, USA\\
$^{\ddagger}$These authors contributed equally to this work.
}

\begin{document}
\label{firstpage}
\pagerange{\pageref{firstpage}--\pageref{lastpage}}
\maketitle

\begin{abstract}

The morphology of a spectral-line profile contains information beyond scalar summaries of line strength, centroid, width, global asymmetry, or diagnostic line ratios. Broad wings, shoulders, double peaks, secondary components, and composite emission--absorption structures encode how flux is ordered across wavelength but can remain indistinguishable under conventional summaries. We introduce an interpretable geometric representation of line profiles inspired by rough path theory. Each wavelength-sampled profile is mapped to a common systemic rest-frame velocity grid and treated as a trajectory in velocity--flux space, traversed from blue to red. From this path, we define a compact set of low-order descriptors measuring signed velocity--flux area, blue--red imbalance localization, higher-order shape complexity, and emission--absorption ordering.
Using synthetic profiles, we show that these descriptors separate morphologies with similar full width at half maximum (FWHM), non-parametric velocity width ($W_{80}$), and low-order moment summaries. We then apply the method to MaNGA integral-field spectroscopy by computing H$\alpha$ descriptors in individual spaxels and clustering them in a low-dimensional feature space. The resulting classes form spatially coherent regions of similar ordered line morphology. Although no external velocity field is supplied to the clustering, stacked spectra within these regions recover coherent large-scale centroid-velocity patterns broadly consistent with the MaNGA reference velocity fields. We release a minimalist MIT-licensed package \texttt{spectropath}, available at \href{https://rafaelsdesouza.com.br/spectropath/}{the project website}.
\end{abstract}

\begin{keywords}
methods: data analysis – methods: statistical – techniques: spectroscopic – galaxies: kinematics and dynamics – software: data analysis
\end{keywords}


\section{Introduction}
\label{sec:introduction}

Scientific progress often depends on finding representations that make the relevant structure of a problem explicit \citep{grattan1997fontana}. Fourier expansions made signals amenable in frequency space \citep{fourier1822}, Bayesian probability provided a calculus for uncertainty \citep{Bayes1763,laplace1820}, Hamiltonian and Lagrangian mechanics reformulated dynamical evolution \citep{Lagrange1788,Hamilton1834}, differential geometry gave a natural language for general relativity \citep{Einstein1916}, and information theory reframed communication in terms of entropy and coding \citep{shannon48}. The common lesson is that a representation is not merely a change of notation, but rather defines which aspects of a system become easy to measure, compare, compress, and interpret.

In astronomy, task-aware compression appears in several related but distinct forms. Spectroscopic and photometric surveys have long been represented through Karhunen--Loève or principal-component projections, which reduce high-dimensional data by retaining directions of dominant variance \citep{Tegmark1997,Connolly1995,ConnollySzalay1999,Yip2004a,Yip2004b}. A more explicitly inferential form of compression is used for galaxy spectra, where optimized linear summaries can be constructed to preserve the Fisher information relevant to a chosen set of physical parameters, rather than merely preserving variance \citep{HeavensJimenezLahav2000,Reichardt2001,Alsing2018}. In the ideal case of parameter-independent noise covariance, a spectrum with thousands of wavelength pixels can be compressed into one number per parameter without loss of Fisher information, making the compressed representation directly tied to model sensitivities such as age, normalization, metallicity, or star-formation history. Other astronomical data products are compressed through sparse and multiscale decompositions, which preserve localized image structure across position and scale \citep{starck2006}, or through shapelet bases, which encode galaxy morphology in a compact set of analytic coefficients \citep{Refregier2003}. Related representations include stellar-population synthesis and spectral decomposition \citep{CidFernandes2005,Sanchez2022}, equivalent widths and Lick-style absorption indices \citep{Worthey1994,Trager2000}, as well as machine-learning methods for denoising and feature extraction \citep{Bu2015,Wang2017}. In this sense, compression should be understood not only as a reduction in data volume, but as a choice of representation. Its value lies in preserving those aspects of the data that are informative for a given scientific question, while suppressing degrees of freedom that are incidental to that question. Here we follow the same guiding principle in the context of spectral-line morphology.

A spectral-line profile contains more than an integrated flux, centroid, width, or global asymmetry: it is an ordered function of wavelength, or equivalently of line-of-sight velocity, in which emission and absorption occur at specific positions relative to the systemic frame. This ordering is physically meaningful because different mechanisms imprint structure in different parts of the profile. Blue-shifted interstellar absorption in NaI D-doublet lines is a classical signature of starburst-driven outflowing material along the line of sight \citep{Heckman2000,Veilleux05}, while broad and blueshifted [O{\sc iii}] components are commonly used to trace ionized outflows in active galaxies \citep{Mullaney2013}. Double-peaked narrow emission lines make this loss of ordered structure especially clear. A two-component profile is not simply a broader line: the relative placement and strength of the blue and red components indicate that the flux is distributed between distinct velocity channels, although similar unresolved profiles may arise from disk rotation, narrow-line-region outflows,  dual active nuclei \citep{Muller2015}, or biconical flows \citep{Nevin2018}, and double-peaked signatures can also be produced by the central rotation curve and bars in disc galaxies \citep{Maschmann2023}. 
Mixed emission--absorption profiles make the role of ordering especially explicit: P-Cygni profiles, with blueshifted absorption and redshifted emission, are associated with outflowing material \citep{Sakamoto2009}, whereas inverse P-Cygni profiles or redshifted absorption are used as evidence for inflowing gas \citep{Herrera2020}. Thus, two profiles with similar flux, width, or scalar asymmetry can still encode different physical configurations if the sequence of blue-side absorption, blue-side excess, core emission, red-side excess, and absorption differs across velocity.
Quantities such as line strength, centroid velocity, velocity dispersion, percentile widths, asymmetry measures, and Gauss--Hermite coefficients provide compact and physically useful summaries of spectra and line-of-sight velocity distributions \citep{vanDerMarelFranx1993,CappellariEmsellem2004}.  Our aim is complementary: to construct descriptors that preserve the ordered morphology of individual line profiles. To achieve this, we will employ path signatures.


Path signatures provide a natural mathematical representation for ordered data. Originating in the theory of iterated integrals and rough paths \citep{Chen1958,Lyons1998}, the signature of a path maps an ordered curve into a hierarchy of tensor coordinates. These coordinates record not only the net displacement of the curve, but also the precise sequence in which its coordinate increments are accumulated. Under suitable regularity conditions, the full signature provides a highly informative description of an ordered curve while remaining invariant under reparameterization \citep{BOEDIHARDJO2016}.
We formalize this by treating a continuum-subtracted line profile not merely as a discrete vector of flux samples
$(F_1,F_2,\ldots,F_n)$,
but as a directed path,
$X(v)=(v,F(v))$, traversed continuously from blue to red velocity in the velocity–flux plane. In this framework, wings, shoulders, and complex emission–absorption structures become geometric features of a directed curve. Because path signature coefficients explicitly record the sequence of variations, a blue-side feature followed by a red-side feature yields a fundamentally distinct representation from the reverse configuration.

While existing software such as \texttt{iisignature} \citep{Reizenstein2020}, \texttt{signatory} \citep{kidger2021signatory}, \texttt{esig} \citep{esig2017}, and \texttt{RoughPy} \citep{Morley2024} provide off-the-shelf implementations of path signatures, the accompanying package developed for this work, \texttt{spectropath}, occupies a more specialized locus in this ecosystem. Instead of acting as a domain-agnostic engine, \texttt{spectropath} fixes the exact astrophysical conventions required for spectral analysis, managing the blue-to-red velocity–flux path mapping, custom normalization routines, and a compact set of low-order descriptors tailored for direct morphological interpretation.

In this work, we introduce these low-order path-signature descriptors to characterize astrophysical line profiles. We show that the coefficients provide interpretable probes of signed velocity--flux area, the localization of asymmetry, core-versus-wing structure, higher-order bends, and the ordering of emission and absorption. Using synthetic profiles, we find that these descriptors separate distinct morphological families most clearly when the relevant difference is encoded in the sequence of flux variations. We then apply the framework to MaNGA integral-field spectroscopy, computing H$\alpha$ path descriptors spaxel by spaxel. Clustering in this geometric space isolates spatially coherent regions with distinct ordered line profiles, whose stacked spectra recover coherent large-scale kinematic structure.

The remainder of this paper is structured as follows. In Section~\ref{sec:background}, we provide an introduction to the path-signature formalism and develop the low-order descriptors used in our analysis. In Section~\ref{sec:applications}, we evaluate their performance in synthetic  line profiles and demonstrate their utility in MaNGA integral-field data. We summarize our results, discuss current limitations, and outline future extensions in Section~\ref{sec:conclusions}.

\section{Background on the path signature}
\label{sec:background}

The path signature is a feature map for ordered, path-shaped data. Unlike summaries based only on pointwise values or global moments, it records how increments accumulate along the path.
In this section we will give a brief introduction to the path signature and its key properties.
We focus on basic examples of piecewise linear paths to convey intuition for the path signature and its behaviour before applying it to real astronomical data in subsequent sections.

\subsection{The path signature and some of its properties}

In this discussion we will assume that the data under study form piecewise smooth curves in $\bbR^n$, i.e.~each data point consists of $n$-many numerical measurements.
A frequently encountered case is where a dataset consists of a string of observations ordered by a single parameter (such as, for instance, frequency or observation times), and where each datapoint is a point in $\bbR^n$ for some fixed $n \in \mathbb{N}$.
Then, such a data stream gives rise to a piecewise smooth curve by connecting the data points in a chosen way (such as linearly interpolating between adjacent data points).

Let $P\bbR^n$ denote the space of piecewise smooth curves in $\bbR^n$.
The path signature for these data can be understood as a map
\begin{equation}
    S \colon P \bbR^n \longrightarrow \bar{T}\bbR^n\,,
\end{equation}
where $\bar{T}\bbR^n$ is the `completed tensor algebra' of $\bbR^n$.
It is the vector space consisting of possibly infinite sums of tensor products of vectors in $\bbR^n$.

As an example, given vectors $w, w_1, w_2$ in $\bbR^n$, each of these vectors defines an element in $\bar{T}\bbR^n$, as do linear combinations like
\begin{equation}
    1 + w_1 + w \otimes w_2\,.
\end{equation}
The difference between the standard tensor algebra and the completed tensor algebra is that the latter also contains infinite sums of tensors.
For example, for each vector $w \in \bbR^n$, there is an element
\begin{equation}
    \exp_\otimes(w) = \sum_{k = 0} \frac{1}{k!} w^{\otimes k}
\end{equation}
of $\bar{T}\bbR^n$.
Here we have used the short-hand notation
\begin{equation}
    w^{\otimes k} = \underbrace{w \otimes \cdots \otimes w}_{\text{$k$-many times}}\,.
\end{equation}

Explicitly, one can define the path signature of a smooth path $\gamma \colon [0,1] \to \bbR^n$ as the infinite sum
\begin{align}
	\label{eq:iterated_integral_signature}
	&S(\gamma) =
	\\
	&1 + \sum_{k = 1}^\infty \int_{1 \geq t_k \geq \cdots \geq t_1 \geq 0} \dot{\gamma}(t_k) \otimes \cdots \otimes \dot{\gamma}(t_1)\, dt_1 \cdots dt_k\,.
    \notag
\end{align}
As an important example, let $w \in \bbR^n$ be a vector.
Then, for a linear path of the form $\gamma[w] \colon [0,1] \to \bbR^n$, $t \mapsto t w$, we obtain the signature
\begin{equation}
\label{eq: signature of linear path}
    S(\gamma[w]) = \exp_\otimes(w)\,.
\end{equation}
Indeed, in this case one computes for the $k$-th term of the signature:
\begin{align}
    &\int_{1 \geq t_k \geq \cdots \geq t_1 \geq 0} \dot{\gamma}(t_k) \otimes \cdots \otimes \dot{\gamma}(t_1)\, dt_1 \cdots dt_k
    \\
    &= \int_{1 \geq t_k \geq \cdots \geq t_1 \geq 0} \underbrace{w \otimes \cdots \otimes w}_{\text{$k$-many times}}\, dt_1 \cdots dt_k
    \notag
    \\
    &= \int_{1 \geq t_k \geq \cdots \geq t_2 \geq 0} w^{\otimes k}\, t_2\, dt_2 \cdots dt_k
    \notag
    \\
    &= \int_{1 \geq t_k \geq \cdots \geq t_3 \geq 0} w^{\otimes k}\, \frac{t_3^2}{2}\, dt_3 \cdots dt_k
    \notag
    \\
    &= \cdots = \frac{w^{\otimes k}}{k!}\,.
    \notag
\end{align}

A fundamental result on the path signature is its multiplicativity, or `functoriality':
if $\gamma_1, \gamma_2 \colon [0,1] \longrightarrow \bbR^n$ are two piecewise smooth paths, we let $\gamma_2 * \gamma_1$ denote their concatenation, i.e.~the piecewise smooth path given by
\begin{equation}
\label{eq: path concatenation}
    (\gamma_2 * \gamma_1)(t) =
    \begin{cases}
        \gamma_1(2t)\,, & 0 \leq t \leq \frac{1}{2}\,,
        \\
        \gamma_2(2(t - \frac{1}{2}))\,, & \frac{1}{2} \leq t \leq 1\,.
    \end{cases}
\end{equation}
Chen's identity \citep{Chen1958} states that the signature of the concatenated path can be computed as the product of the individual path signatures,
\begin{equation}
    S(\gamma_2 * \gamma_1) = S(\gamma_2) \otimes S(\gamma_1)\,.
\end{equation}
This mathematical property makes the signature extremely well-behaved and can be exploited in computations: it allows us to split paths into time segments, compute the segments' individual signatures in a parallelized fashion, and then take the tensor products of the results (a simple algebraic operation) in order to obtain the signature of the entire path.
One can also check that the path signature $S(\gamma)$ of a path $\gamma$ is independent of the parameterization of the path, as well as translations of $\gamma$ by any constant vector.
Appendix~\ref{app:gauge} provides an additional geometric perspective on these properties by interpreting the signature in terms of parallel transport. This viewpoint also clarifies the functoriality of the signature and several related structural features.
However, that presentation is more involved mathematically, and so we defer it to an appendix for readers interested in a deeper perspective on the signature.

A \textit{piecewise linear path} in $\bbR^n$ is a path which can be written as a concatenation of paths of the form $\gamma[w_i] \colon [0,1] \to \bbR^n$, $t \mapsto t w_i$, for vectors $w_1, \ldots, w_N \in \bbR^n$.
In particular, the linear interpolation of any data stream results in a piecewise linear path.
By the multiplicativity property of the signature and Equation~\eqref{eq: signature of linear path}, the path signature of a piecewise linear path reads as
\begin{align}
    S \big( \gamma[w_N] * \cdots * \gamma[w_1] \big)
    &= S \big( \gamma[w_N] \big) \otimes \cdots \otimes S \big( \gamma[w_1] \big)
    \\
    &= \exp_\otimes (w_N) \otimes \cdots \otimes \exp_\otimes (w_1)\,.
    \notag
\end{align}
We thus obtain the following interpretation of the path signature for piecewise linear paths:
if we write the vector $w_i = (w_i)^\mu e_\mu$ in the standard basis $e_1, \ldots e_n$ of $\bbR^n$, we can think of the vector $(w_i)^\mu \partial_\mu$ as the generator of an infinitesimal translation along $w_i$.
The actual, finite translation established by the path $\gamma[v_i]$ can then be understood as the family of translations
\begin{equation}
    t \mapsto \exp \big( t\, (w_i)^\mu \partial_\mu \big)\,.
\end{equation}
The path signature $S(\gamma[w_i])$ is now merely a different way to write down the same expression, essentially differing only by replacing $\partial_\mu$ by $e_\mu$ and the product of derivatives by the tensor product.
The signature of the concatenated piecewise linear path $\gamma[w_N] * \cdots * \gamma[w_1]$ is, thus, merely the sequence of finite translations one needs to apply to the origin $0 \in \bbR^n$ in order to trace out this path.
Since any piecewise smooth path can be approximated by piecewise linear ones, we also obtain a similar interpretation of the general path signature as a limit of series of exponentiated generators of translations in $\bbR^n$.

\subsection{The path signature as a feature map}

One key set of properties of the path signature is that it is invariant under reparameterizations, translations, as well as thin homotopies of paths.
Whilst reparameterizations and translations are standard notions, thin homotopy of paths deserves further explanation.
A homotopy of paths is a continuous deformation of one path into another.
Equivalently, it is a 1-parameter family of paths.
This can always be written as a map
\begin{equation}
    h \colon [0,1]^2 \to \bbR^n\,,
    \qquad
    (s,t) \mapsto h(s,t)
\end{equation}
depending on two parameters $s$ and $t$, where the first parameter $s$ controls the position in the 1-parameter family of paths, and the second parameter $t$ controls the location along the path.
A homotopy is called thin if, at every point $(s,t) \in [0, 1]^2$ in the square, the two vectors $\partial_s h(s,t)$ and $\partial_t h(s,t)$ are linearly dependent.
In particular, it can be shown that any such thin homotopy of paths corresponds to contracting a piece of the path where the path moves in a certain direction and then traces back the same piece of path in the opposite direction (i.e.~the path goes back on itself along exactly the same trajectory).

The power of the path signature as a feature map for path-shaped data stems from the fact that, under standard regularity assumptions, the full signature characterizes a path up to translation, reparameterization, and tree-like equivalence, a notion closely related to thin homotopy \citep{HamblyLyons2010,BOEDIHARDJO2016}.

That is, the full, untruncated signature really remembers all the information contained in the input path, up to reparameterization, translation and thin homotopy. Even more, when viewed as a map from the space of all paths in $\bbR^n$ to the completed tensor algebra, it can even distinguish a large class of probability measures on the space of all paths, even in the case where paths are much less well-controlled than the piecewise smooth paths we have been discussing here \citep{ChevyrevLyons2016}. Thus signatures are not only descriptors of individual paths, but can also serve as feature maps for distributions of paths.

In order to deploy the path signature as a feature map, several additional observations are important:
first, the path signature as defined above transforms potentially infinite-dimensional data points (piecewise smooth paths in $\bbR^n$) into potentially infinite-dimensional data points (vectors in the completed tensor algebra $\bar{T}\bbR^n$).
While the full infinite-dimensional information is important from a theoretical perspective in order to prove the range of strong properties the signature enjoys, for practical purposes infinite-dimensional data points are naturally infeasible.

Instead, one may think of the path signature in a way similar to Taylor expansions or the Fourier transform.
Relevant information about the data at hand is already contained in low orders of the signature, and will approximate the full data increasingly well if increasingly high orders are taken into account.
There is, however, a balance to strike between computational cost and accuracy.
In practical applications of the signature, one thus truncates the output of the signature in a certain tensor order, which needs to be specified.

For example, for a linear path, the signature truncated in order $k$ reads as
\begin{equation}
    S_{\leq k}(\gamma[w]) = \sum_{i = 0}^k \frac{w^{\otimes i}}{i!}\,.
\end{equation}
For consistency, when multiplying $k$-truncated path signatures, we use the standard product in $\bar{T}\bbR^n$, but discard all terms containing more than $k$ tensor factors.

Second, depending on the concrete application, the signature's invariance under thin homotopies and reparameterization can either be a very strong feature which naturally discards distortions of the data, or it can be an obstacle, for instance if it is important to record the distance between data points in the variable parameterizing the paths. Below this will be, for example, the wavelength in IFU spectra.

In cases which fall in the second bracket, there is a straightforward fix.
One passes from the signature $S(\gamma)$ of a path $\gamma \colon [0, 1] \to \bbR^n$, $t \mapsto \gamma(t)$, to its time-augmented signature, which is the signature $S(\tilde{\gamma})$ of the path $\tilde{\gamma} \colon [0,1] \to \bbR^{n+1}$, $t \mapsto (t, \gamma(t))$.
That is, the time-augmented path $\tilde{\gamma}$ records not only the value of the observations (like the path $\gamma$), but also keeps track of the parameter-value at which each observation in a path-shaped data set has been recorded.

\begin{figure}
    \centering
    \includegraphics[width=0.9\linewidth]{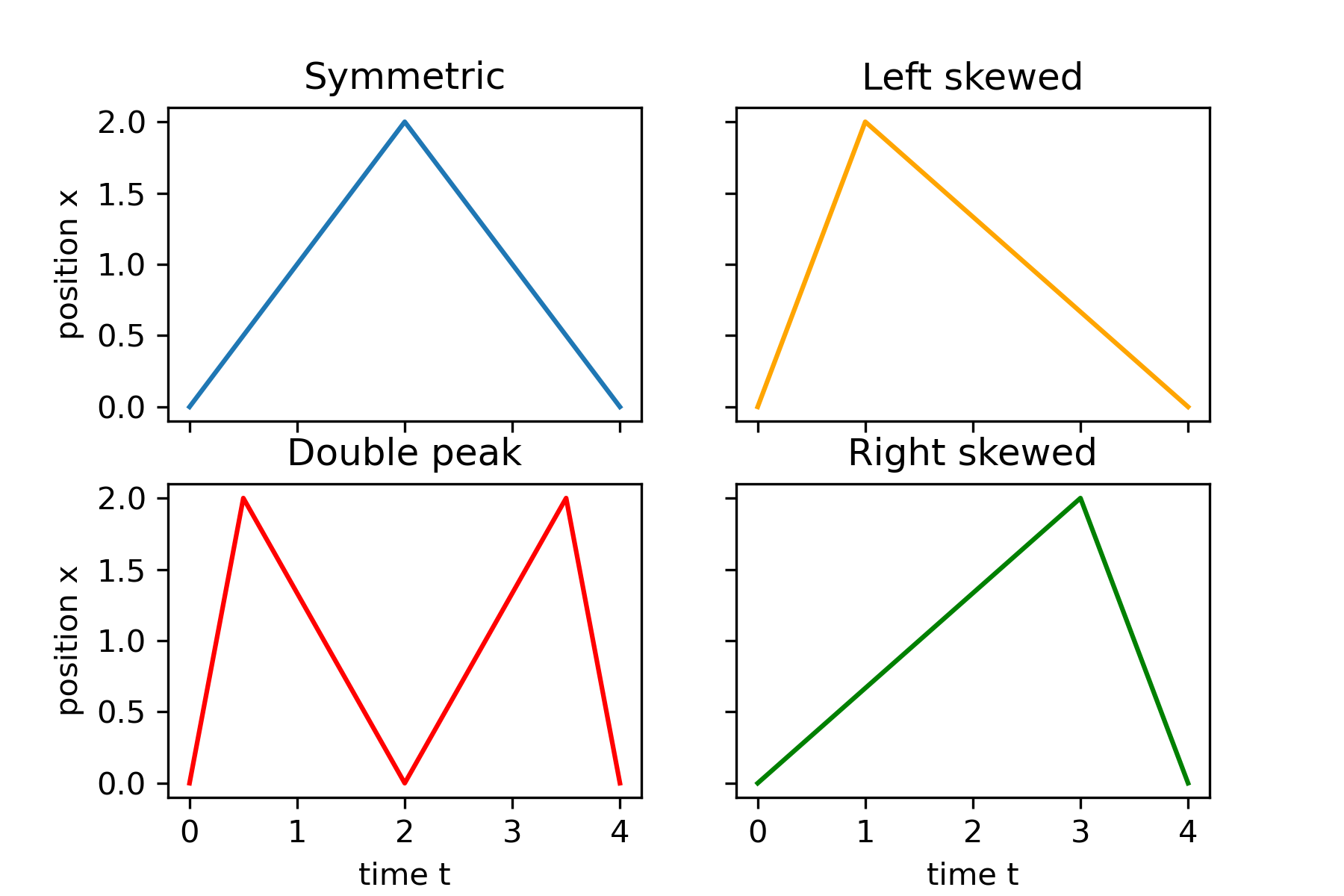}
    \caption{Four piecewise linear paths in one dimension ($y$-axis), augmented by the time component ($x$-axis).}
    \label{fig: PL signatures example}
\end{figure}

In the following, we will mainly combine this augmentation with the idea of truncating the signature for computational feasibility.
We finish this subsection by considering the four paths depicted in Figure~\ref{fig: PL signatures example} as an example.
The figure shows time-augmented versions of four piecewise linear paths in one dimension (the augmentation adds a second component).
One can alternatively think of these graphs (up to rescaling) as toy models for probability distributions or measurements of spectra.
The time-augmented signature of the underlying paths in one-dimension is the same as the signature of the paths in two dimensions shown in the figure.

We computed the signature of these piecewise linear paths up to tensor order $N = 4$.
Up to order two, the signatures of all paths in Figure~\ref{fig: PL signatures example} agree.
(The zeroth order is always equal to the number $1 \in (\bbR^n)^{\otimes 0} = \bbR$.)
The first order signature equals the total displacement, i.e.~the difference $\gamma(1) - \gamma(0)$.
These displacements coincide for all four paths shown in the figure.
The signatures of the symmetric and skewed paths differ at order $3$, as do the signatures of the left and right skewed paths.
The signatures of the symmetric and double-peak paths differ at order $4$.
This provides some intuition for which features of path-shaped data different orders of the path signature distinguish. We will comment more on this question below, as well as provide a much extended and more realistic range of experiments that we hope will help illustrate which features of curves selected components of the signature detect.

\subsection{The log-signature}

The tensor coordinates of the signature are highly structured and satisfy algebraic relations, such as shuffle identities. At the same time, this means that there is redundancy in the collection of components of the path signature. It is therefore often useful
to pass from the signature to its logarithm in the completed tensor algebra.
The log signature of a path $\gamma$ can be defined from its signature by applying the map
\begin{equation}
    \log \colon \bar{T} \bbR^n \to \bar{T} \bbR^n\,,
    \qquad
    x \mapsto \sum_{m = 1}^\infty \frac{(-1)^{m+1}}{m} (x - 1)^m\,,
\end{equation}
amounting to the Taylor expansion of the logarithm function.
Concretely, substituting for $x$ the signature $S(\gamma)$ of a path $\gamma$, the log signature of $\gamma$ reads as
\begin{equation}
    L(\gamma)=\log S(\gamma)
    =
    \sum_{m=1}^{\infty}
    \frac{(-1)^{m+1}}{m}
    \left(S(\gamma)-1\right)^m .
\end{equation}
As an example, the log signature of a linear path reads as
\begin{equation}
    L(\gamma[w]) = w\,.
\end{equation}

The components of the log signature coordinates live in the free Lie algebra
generated by the coordinate directions of the path, i.e.~the free Lie algebra on $\bbR^n$, and provide a non-redundant description of the same ordered geometry.
The multiplicativity of the path signature now translates to the log signatures of paths multiplying via the Baker-Campbell-Hausdorff formula in the free Lie algebra on $\bbR^n$.

At first order, the log-signature records endpoint displacement. In the emission-line setting considered here, each line profile is represented as a two-dimensional path in velocity--flux space, \((v,F(v))\) (where we denote the original path by $F$ and its parameter by $v$, so that the `time-augmented path' reads as $(v, F(v))$). In the concrete applications below, the coordinate
\(v\) denotes the Doppler velocity offset from the systemic rest-frame line centre, while \(F\) denotes the continuum-subtracted line flux, or a normalized version of this flux when only the profile shape is used. Thus, first-order
terms summarize the net displacement of the profile across velocity--flux space. At second order, the log-signature records antisymmetric area terms, which measure the oriented area swept out by the profile and are therefore
sensitive to line-profile asymmetries, such as blue or red wings, skewed emission, and multiple kinematic components. Higher-order terms encode how these oriented areas are distributed and modulated along the path.
We denote log-signature coordinates by superscripts. Thus, for a
two-dimensional path with coordinates \(v\) and \(F\), the terms \(L^{vF}\) and \(L^{Fv}\) represent the two orderings of velocity and flux increments. In the
applications below we do not use the full log-signature. Instead, we select a small number of low-order scalar contrasts whose geometric interpretation is simple for two-dimensional paths of the form \((v,F(v))\).

\section{Astronomical applications}
\label{sec:applications}

The path-signature formalism introduced in Section~\ref{sec:background}
can be applied to many ordered astronomical data sets, including spectra and light curves. In this work, we focus on continuum-subtracted spectra, and in particular on the morphology of spectral-line profiles; however, the same formalism can be extended directly to other ordered astronomical data.

\subsection{Path coefficients for line profiles}
\label{sec:spectral_paths}

We now specialize the abstract path \(\gamma\) of
Section~\ref{sec:background} to a spectroscopic setting. A
continuum-subtracted line profile is represented as the ordered curve
\begin{equation}
    X(u)=\bigl(u,F(u)\bigr),
\end{equation}
where \(u\) is a spectral coordinate, such as wavelength, log-wavelength,
or velocity, and \(F(u)\) is the continuum-subtracted flux. For kinematic
applications we set \(u=v\), with \(v\) measured relative to the adopted line centre, so that the profile is traced from blue to red velocity.
Observed spectra are sampled at discrete pixels, so we treat each profile as
the piecewise-linear path obtained by joining neighbouring samples. In
practice, the profiles are placed on a common velocity scale and normalized
before the coefficients are compared, so that the resulting descriptors are
driven primarily by line morphology rather than by differences in sampling,
velocity range, or overall flux scale. We then compute selected low-order log-signature coordinates of this path and combine them into a small set of
interpretable scalar coefficients. The notation is chosen to be mnemonic: the number gives the order of the log-signature term, while the letter indicates the main geometric interpretation.

At second-order we use the coefficient
\begin{equation}
    p_2
    =
    \frac{1}{2}
    \left(
        L^{vF}-L^{Fv}
    \right).
\end{equation}
Geometrically, \(p_2\) is the antisymmetric second-order
velocity--flux term, equivalent to the signed area swept by the directed
profile in the velocity--flux plane. Here \(v\) is not a fitted gas velocity,
but the common velocity coordinate of the line window after shifting the
spectrum to the systemic rest frame of the galaxy. Thus, for a profile traced
from blue to red velocity, \(p_2\) summarizes the handedness of the
continuum-subtracted curve \((v,F(v))\). It is sensitive to where flux is
distributed relative to the adopted rest-frame line centre, and can therefore
respond to centroid shifts, blue--red imbalance, and net line area. For this
reason we do not interpret \(p_2\) as a complete asymmetry or velocity
diagnostic on its own; it is used together with higher-order coefficients that
encode how the velocity--flux geometry is distributed along the profile.

At third order we use the coefficients
\begin{align}
    p_{3v}
    &=
    \frac{1}{2}
    \left(
        L^{vvF}-L^{vFv}
    \right),
    \\
    p_{3F}
    &=
    \frac{1}{2}
    \left(
        L^{FvF}-L^{FFv}
    \right).
\end{align}
The coefficients \(p_{3v}\) and \(p_{3F}\) are third-order log-signature coordinates built from nested commutators involving the second-order antisymmetric velocity--flux area term. In the free-Lie-algebra description of the log-signature, such terms may be viewed as higher-order modulations of the signed area. In the spectroscopic setting, we therefore interpret \(p_{3v}\) heuristically as emphasizing where profile asymmetry accumulates along the velocity axis, and \(p_{3F}\) as emphasizing whether that asymmetry is associated with brighter core emission or fainter wings and shoulders. This is our own empirical interpretation rather than a standalone theorem, and we convey further intuition in the synthetic-profile tests below.

At fourth order we use
\begin{align}
    p_{4F}
    &=
    \frac{1}{4}
    \left(
        L^{FFFv}
        -
        L^{FFvF}
        +
        L^{FvFv}
        -
        L^{vFFF}
    \right),
    \\
    p_{4T}
    &=
    \frac{1}{4}
    \left(
        L^{vFvF}
        -
        L^{vFFv}
        +
        L^{FvvF}
        -
        L^{FFvv}
    \right).
\end{align}
The coefficient \(p_{4F}\) responds to higher-order flux structure,
such as shoulders, reversals, and broad bases. The coefficient \(p_{4T}\)
responds to twist-like velocity--flux structure, including bends, double peaks and multiple components.
For profiles containing both emission and absorption, we additionally define
an emission--absorption ordering coefficient. This is not computed directly
from the original velocity--flux path, but from the cumulative positive and
negative parts of the profile. We write
\begin{align}
    C_+(v)
    &=
    \int_{v_{\rm min}}^{v}
    \max\{F(v'),0\}\,dv',
    \\
    C_-(v)
    &=
    \int_{v_{\rm min}}^{v}
    \max\{-F(v'),0\}\,dv' .
\end{align}
The corresponding signed area is the L\'evy area of the cumulative
emission--absorption path \(Y(v)=(C_+(v),C_-(v))\):
\begin{equation}
    p_{\pm}
    =
    \frac{1}{2}
    \int_{v_{\min}}^{v_{\max}}
    \left[
        C_+(v)\,dC_-(v)
        -
        C_-(v)\,dC_+(v)
    \right].
\end{equation}
Thus \(p_{\pm}\) measures the oriented area swept out in the
\((C_+,C_-)\) plane as one moves from blue to red velocity. It is therefore sensitive to whether absorption tends to occur before emission, or emission before absorption, along the ordered profile. This distinguishes, for example, P-Cygni-like profiles from inverse P-Cygni-like profiles.

\begin{table}
\centering
\caption{
Qualitative interpretation of the path coefficients used in this work.
}
\label{tab:path_coefficients}
\renewcommand{\arraystretch}{1.2}
\setlength{\tabcolsep}{5pt}
\begin{tabular}{lp{5.8cm}}
\toprule
\textbf{Coefficient} & \textbf{Main interpretation} \\
\midrule
\(p_2\)
& Signed velocity–flux area. \\

\(p_{3v}\)
& Velocity location of profile imbalance (center vs wings). \\

\(p_{3F}\)
& Flux-weighted imbalance. \\

\(p_{4F}\)
& Higher-order flux modulation (shoulders and broad bases). \\

\(p_{4T}\)
& Twist-like velocity--flux structure (double peaks, and multiple components). \\

\(p_{\pm}\)
& Relative ordering of emission and absorption in mixed-sign profiles. \\
\bottomrule
\end{tabular}
\end{table}

\subsection{Synthetic spectral-line profiles}
\label{sec:mock_profiles}

\begin{figure*}
\centering
\includegraphics[width=\linewidth]{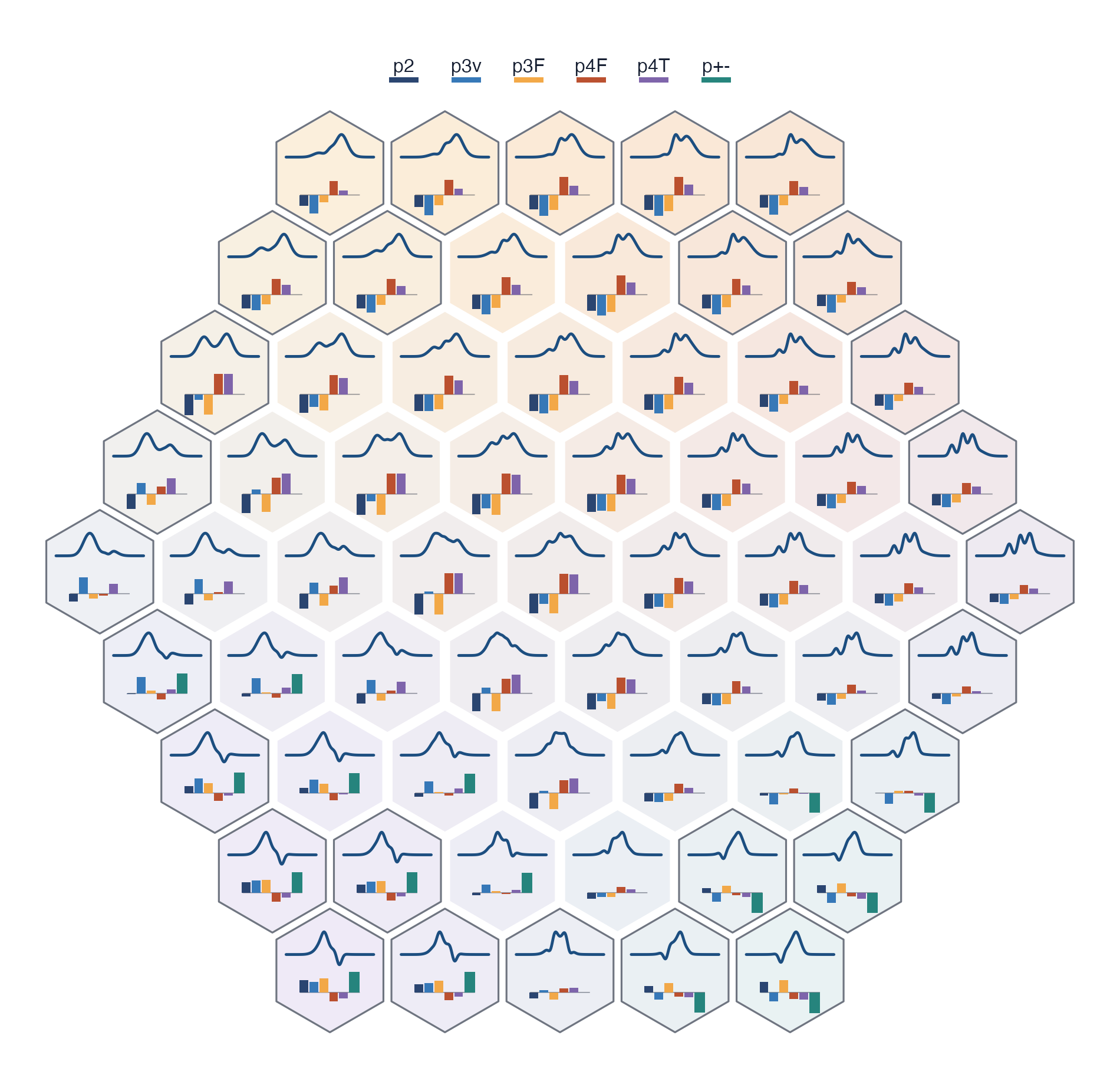}
\caption{
Schematic response atlas for low-order path coefficients. Each hexagonal cell
shows an idealized continuum-subtracted line profile traced from blue to red
velocity. The atlas interpolates between six profile archetypes, including
shifted single-component lines, wing or shoulder structure, multi-component
profiles, and the two opposite orderings of emission--absorption structure.
The small bars below each profile show signed path-coordinate responses
relative to a symmetric reference profile; bars above and below the local
zero baseline indicate opposite signs of the same coordinate. The coefficients
\(p_2\) and \(p_{3v}\) are sensitive to directed velocity--flux imbalance,
\(p_{3F}\) and \(p_{4F}\) to flux-weighted redistribution, shoulders, and
broad bases, \(p_{4T}\) to higher-order multi-component structure, and
\(p_{\pm}\) to the ordering of emission and absorption. The atlas is intended
as a qualitative guide to path-coordinate sensitivity rather than a physical
calibration.
}
\label{fig:path_profile_manifold}
\end{figure*}

To build intuition for these coefficients, we visualize their responses on a
continuous family of idealized line profiles. The purpose of this
construction is not to calibrate the coefficients against physical parameters,
but to show how different ordered morphologies activate different components
of the path vector. The atlas is organized around six archetypal profile
directions: shifted single-component lines, wing or shoulder structure,
multi-component profiles, and the two opposite orderings of
emission--absorption structure. Each profile is evaluated with the same path
feature definitions used elsewhere in the paper. The small bars below each
profile show signed responses relative to a symmetric reference line, so bars
above and below the zero baseline indicate opposite orientations of the same
coordinate. This is particularly important for \(p_{\pm}\), which responds to
the ordering of positive and negative flux and therefore changes sign between
P-Cygni-like and inverse-P-Cygni-like profiles. The resulting atlas
(Fig.~\ref{fig:path_profile_manifold}) should be read as a qualitative map of
path-coordinate sensitivity: changes along the velocity axis, redistribution
of flux between core and wings, shoulders or multiple components, and
emission--absorption ordering produce different low-order path-response
patterns. This motivates using selected path coordinates as compact,
morphology-aware descriptors of emission-line profiles.

\begin{figure*}
\centering
\includegraphics[width=\linewidth]{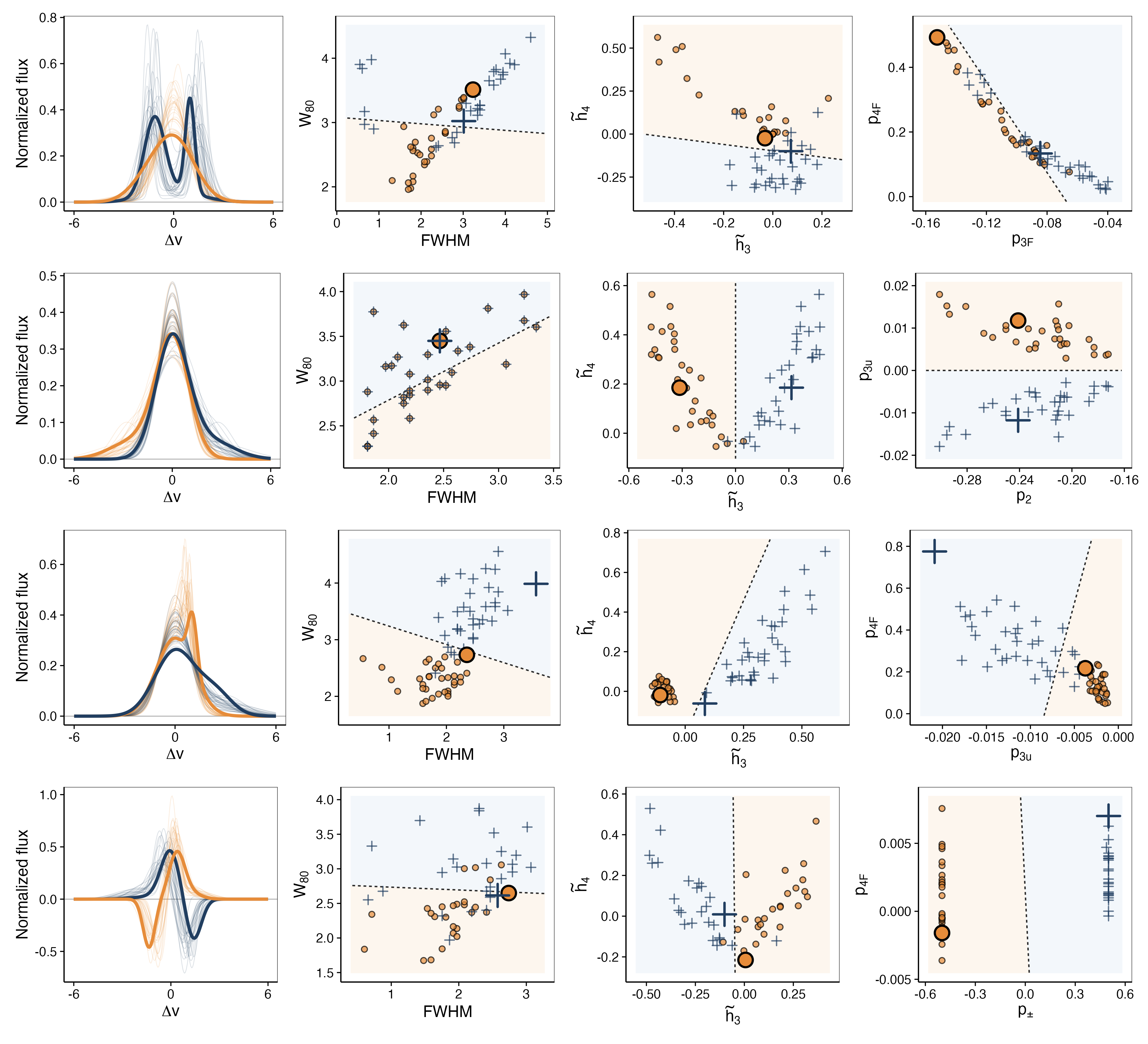}
\caption{Toy line-profile families used to compare classical and path-based morphology
descriptors. In each row, the left panel shows two representative profiles in
velocity-offset coordinates, with faint curves showing nearby profiles drawn
from the same synthetic family. The middle panels show the corresponding
locations in conventional diagnostic spaces based on FWHM, \(W_{80}\), and
Hermite-like summaries \(\tilde h_3\) and \(\tilde h_4\). The right panel
shows a path-coordinate plane selected for that morphology family. Enlarged
symbols mark the representative profiles displayed on the left. Shaded regions
show a simple linear separation in each diagnostic plane and are included only
as a visual guide. The examples illustrate cases in which profiles overlap
under width- or moment-based summaries but separate more clearly when the
ordered geometry of the line profile is retained.}
\label{fig:family_profiles}
\end{figure*}

The response atlas shows how individual path coordinates behave on a smooth,
idealized morphology manifold. We next ask a complementary question: whether
these coordinates can improve the separation of profile families when compared
with commonly used scalar summaries. For this purpose we construct synthetic line-profile experiments in which two families are generated with
overlapping widths or global moments but different ordered structure. These
experiments are not intended as realistic forward models of any particular
galaxy or ionized-gas component. Instead, they isolate profile-shape
contrasts, such as blue versus red wings, single versus double components,
core--wing redistribution, and emission--absorption ordering, and compare how
well different diagnostic spaces preserve those contrasts.

For each synthetic profile we define the ordered curve
\begin{equation}
X(\Delta v) = \bigl(\Delta v, F(\Delta v)\bigr),
\end{equation}
where \(\Delta v\) is the velocity offset from the adopted line centre and
\(F\) is the normalized continuum-subtracted flux. We compute the path
coefficients listed in Table~\ref{tab:path_coefficients} and compare them
with conventional summaries including FWHM, \(W_{80}\), and Hermite-like
shape summaries \(\tilde h_3\) and \(\tilde h_4\).

Figure~\ref{fig:family_profiles} shows representative examples. The
FWHM--\(W_{80}\) plane is primarily sensitive to the overall velocity extent
of the line, and is therefore useful for describing broadening, but it can be
insensitive to the direction or ordering of profile structure. The
Hermite-like plane captures global skewness and peakedness, but still reduces
the profile to distributional shape summaries. By contrast, the path-coordinate
planes depend on how the profile is traced from blue to red velocity. Profiles
that remain close in width- or moment-based spaces can therefore separate in
path space when their ordered morphology differs, for example between blue and
red wings, single and double peaks, core--wing redistribution, or
P-Cygni-like and inverse-P-Cygni-like ordering.

To quantify the visual separability in Fig.~\ref{fig:family_profiles}, we use
the full set of synthetic profiles generated for each morphology experiment, rather
than only the two highlighted examples. For each diagnostic plane \(P\), the
two coordinates are robustly standardized within that experiment by subtracting
the median and dividing by the median absolute deviation. We then compute a
simple linear score
\begin{equation}
s_i^{(P)} =
\boldsymbol w^{(P)}\cdot \tilde{\boldsymbol z}_i^{(P)},
\end{equation}
where \(\tilde{\boldsymbol z}_i^{(P)}\) is the standardized coordinate of
profile \(i\), and \(\boldsymbol w^{(P)}\) is the direction that best separates
the two synthetic-profile families in that plane. The shaded regions in
Fig.~\ref{fig:family_profiles} show this linear separation and are intended
only as a visual guide.

\begin{figure*}
\centering
\includegraphics[width=\linewidth]{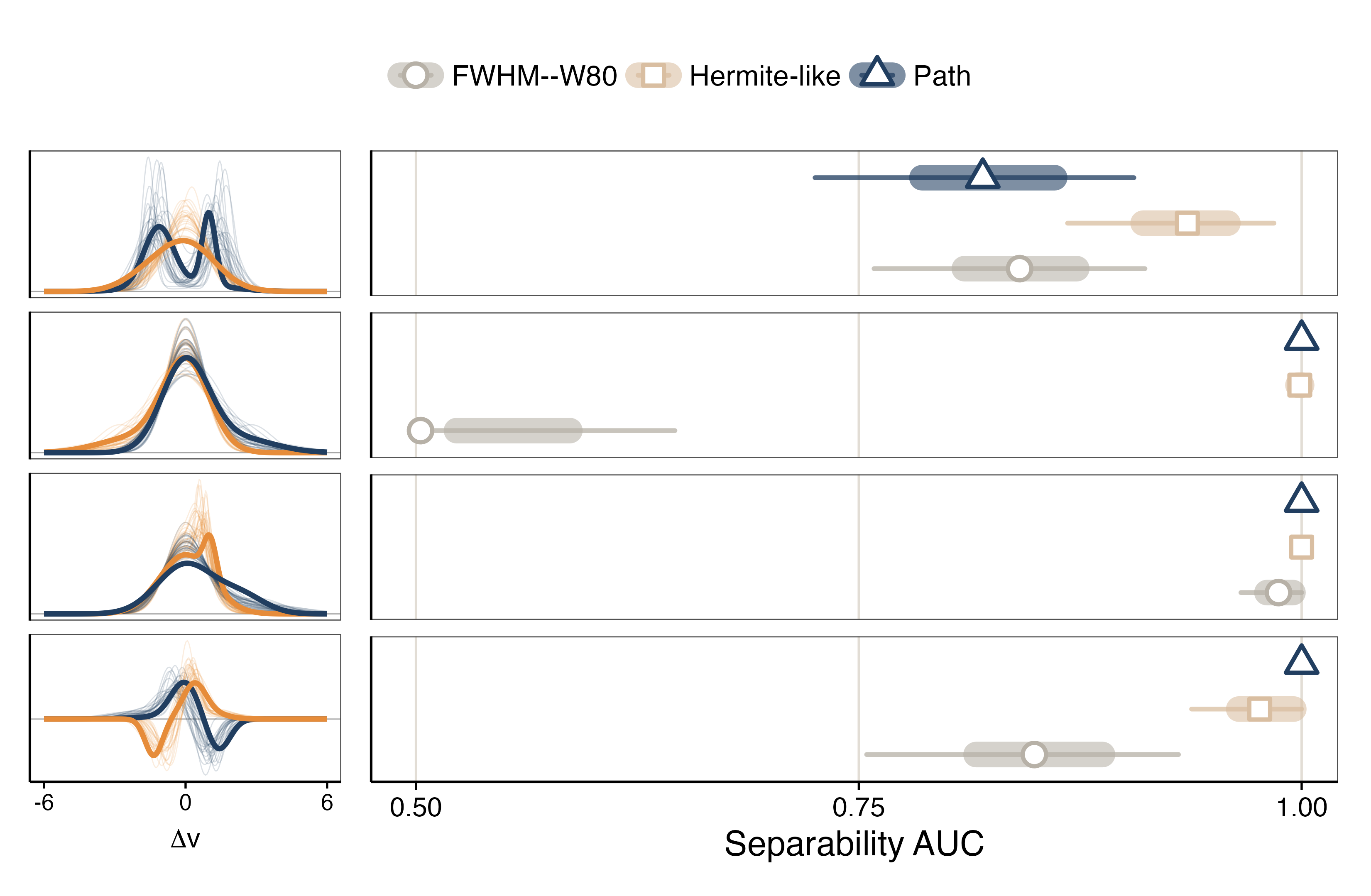}
\caption{
Family-level separability of the synthetic profile experiments. The small panels on
the left show the two profile families in each row, including nearby synthetic
profiles generated by varying the same profile parameters. The intervals on
the right show bootstrap estimates of the separability AUC for three
diagnostic planes: FWHM--\(W_{80}\), the Hermite-like
\(\tilde h_3\)--\(\tilde h_4\) plane, and the selected path-coordinate plane.
 Path coordinates perform best when the distinction is
controlled by ordered morphology, such as blue--red ordering, core--wing
redistribution, or emission--absorption ordering.
}
\label{fig:toy_quantification}
\end{figure*}

We summarize the separation using the area under the receiver-operating
characteristic curve, \({\rm AUC}^{(P)}\). Equivalently, the AUC is the
probability that a randomly chosen profile from one family is ranked above a
randomly chosen profile from the other family by the score \(s_i^{(P)}\), up
to the arbitrary choice of sign:
\begin{equation}
{\rm AUC}^{(P)}
=
\max\left[
{\rm Pr}\left(s_a^{(P)} > s_b^{(P)}\right),
{\rm Pr}\left(s_b^{(P)} > s_a^{(P)}\right)
\right],
\end{equation}
where \(a\) and \(b\) are drawn from the two distribution families. Thus
\({\rm AUC}=0.5\) corresponds to no separability, while \({\rm AUC}=1\)
corresponds to perfect separation in that diagnostic plane. The intervals in
Fig.~\ref{fig:toy_quantification} are obtained by bootstrap resampling the
synthetic profiles within each family.

This quantity is not used as a classifier for real data. It is a descriptive
score for the controlled experiments: it asks how well a given two-dimensional
representation separates two known morphology families. In this sense,
Fig.~\ref{fig:toy_quantification} provides a compact quantitative counterpart
to the visual comparison in Fig.~\ref{fig:family_profiles}.

As a complementary stress test, we repeated the controlled experiments using profiles constructed from mixtures of Moffat-like components. The Moffat family provides a tunable transition between a sharply concentrated core and extended wings through its shape parameter, allowing more varied core--wing structure than the Gaussian components used in the preceding examples. We used these components to generate families with shoulders, secondary peaks, asymmetric wings, and signed emission--absorption structure, while retaining the same diagnostic and quantification procedure described above. This experiment tests whether the relative behaviour of the diagnostic spaces depends strongly on the adopted analytic profile family.

\begin{figure*}
\centering
\includegraphics[width=\linewidth]{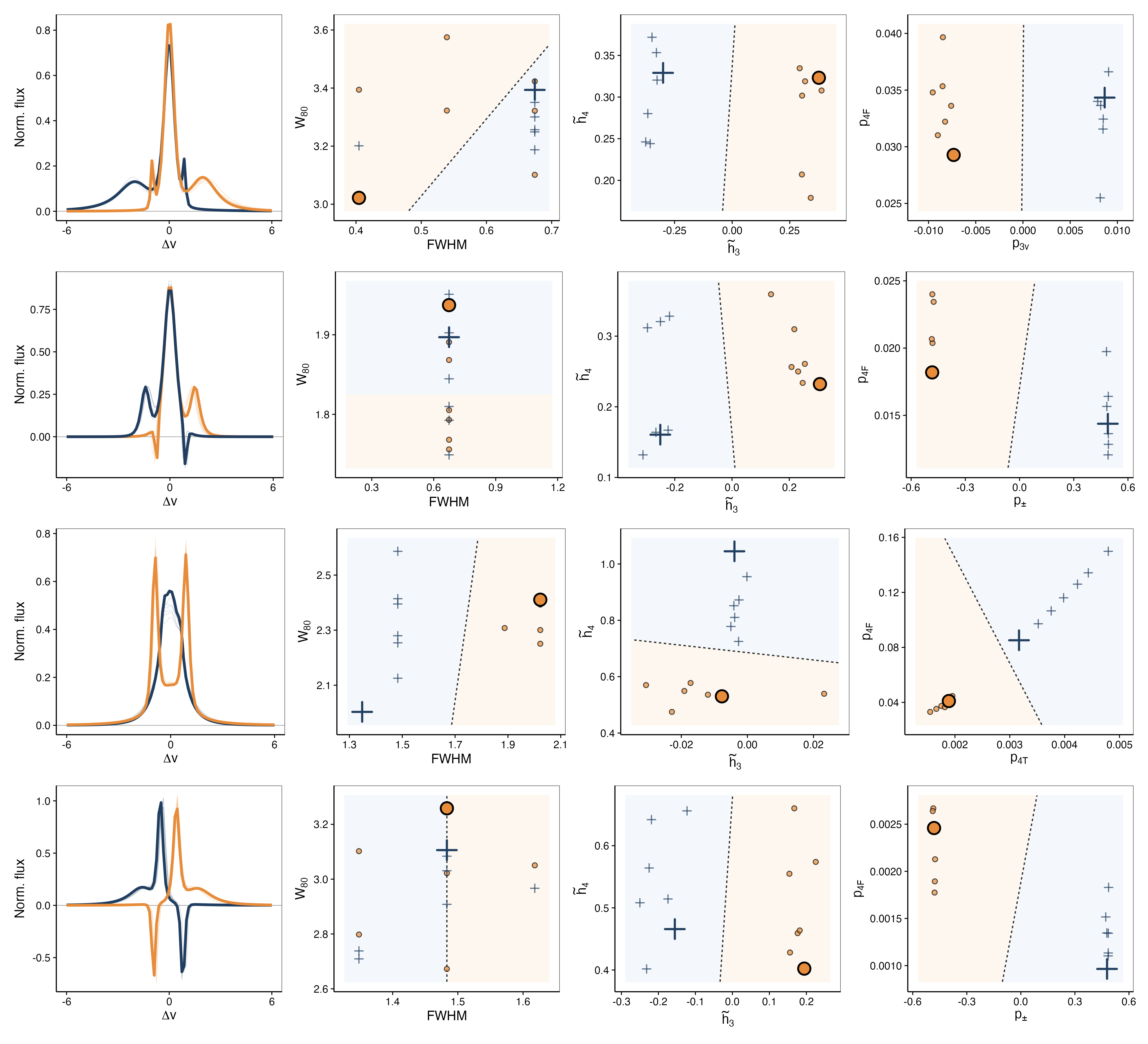}

\caption{
Stress tests based on mixtures of Moffat-like line components. Each row compares
two synthetic-profile families generated by varying components with tunable core
concentration and extended wings. The left panels show representative profiles
from the two families, while the remaining panels show the corresponding
ensembles in the $\mathrm{FWHM}$--$W_{80}$, $\tilde h_3$--$\tilde h_4$, and
selected path-coordinate planes. Enlarged symbols mark the archetypal profiles;
smaller points show samples drawn from each family distribution. The shaded
regions and linear boundaries are included only as visual guides. Width-based
coordinates provide limited discrimination when the families have similar
velocity extent, whereas path coordinates distinguish profiles that differ in
the placement or blue-to-red ordering of shoulders, wings, secondary components,
or emission--absorption structure. Hermite-like coordinates remain effective
when the families also differ in global asymmetry or core--wing concentration.
}
\label{fig:moffattt}
\end{figure*}

Figure~\ref{fig:moffattt} shows that the path coordinates continue to provide information beyond the conventional width measures. In particular, the FWHM--$W_{80}$ plane often overlaps substantially when the families have comparable velocity extent but differ in the placement or ordering of their substructure. The comparison with the Hermite-like plane is more stringent. Because Moffat mixtures can also differ in global asymmetry and core--wing concentration, \(\tilde h_3\) and \(\tilde h_4\) remain sensitive to several of the induced variations and are therefore competitive in some experiments.
The path coordinates are most useful when the separation depends less on overall skewness or peakedness and more on the ordered sequence of features across velocity. This stress test therefore should not be read as evidence for the uniform superiority of one representation. Rather, it shows that the path-based description remains informative for a broader and more flexible family of line-profile morphologies, especially when the discriminating signal lies in the ordering and placement of substructure.

\begin{figure*}
\centering
\includegraphics[width=\linewidth]{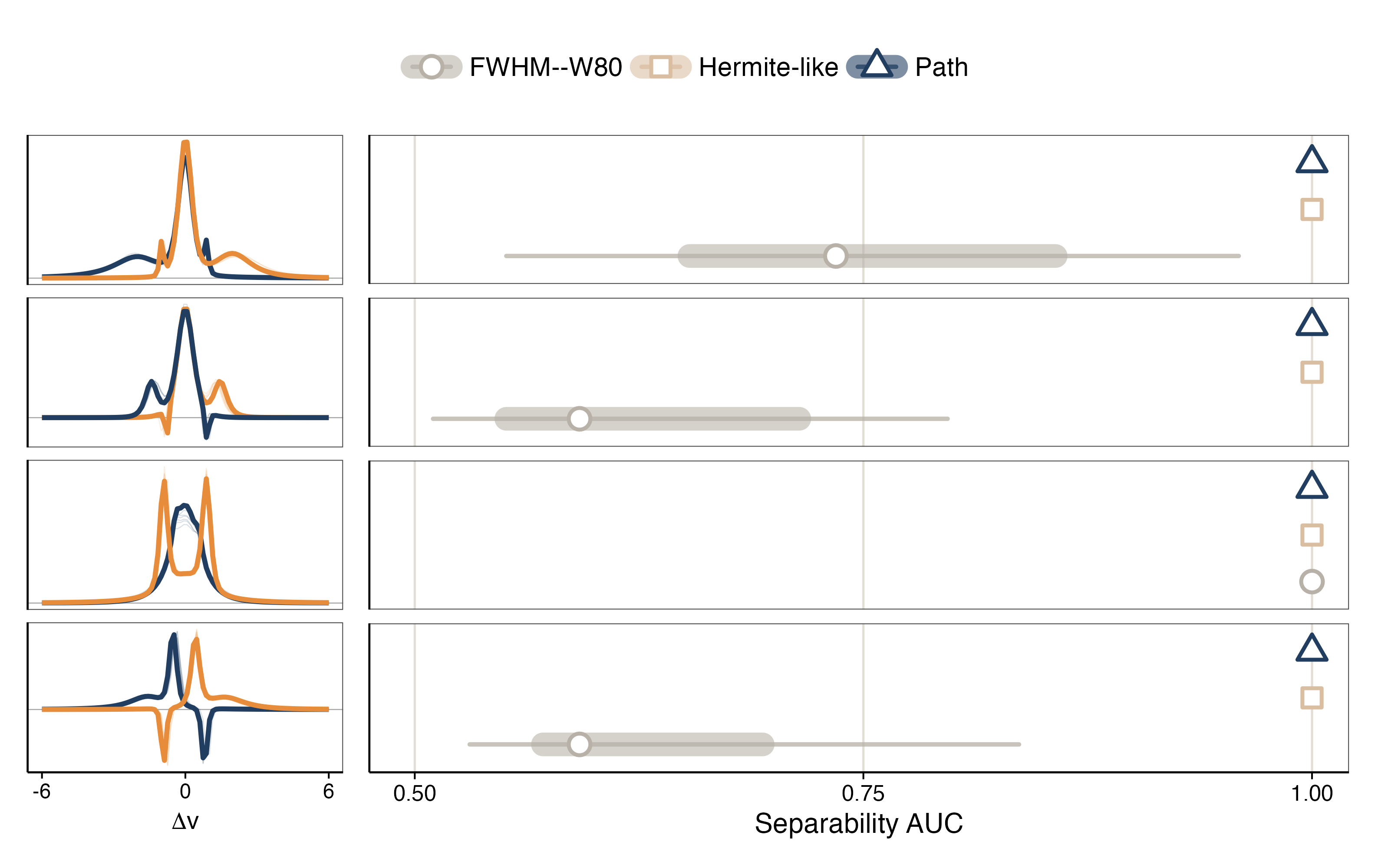}

\caption{
Quantification of the Moffat-mixture stress tests shown in Fig.~\ref{fig:moffattt}. The small panels on the left show the two synthetic-profile families considered in each experiment. The intervals on the right give bootstrap estimates of the separability AUC in the $\mathrm{FWHM}$--$W_{80}$, $\tilde h_3$--$\tilde h_4$, and selected path-coordinate planes.  The path-coordinate representation is most effective when the distinction is governed by the location and ordering of profile substructure rather than solely by changes in width, skewness, or core--wing concentration. The competitive performance of the $h_3$--$\tilde h_4$ plane in some experiments indicates that global shape summaries remain informative when the Moffat mixtures differ appreciably in asymmetry or peakedness.
}
\label{fig:moffat_stress_auc}
\end{figure*}

\begin{figure*}
\centering
\includegraphics[width=\linewidth]{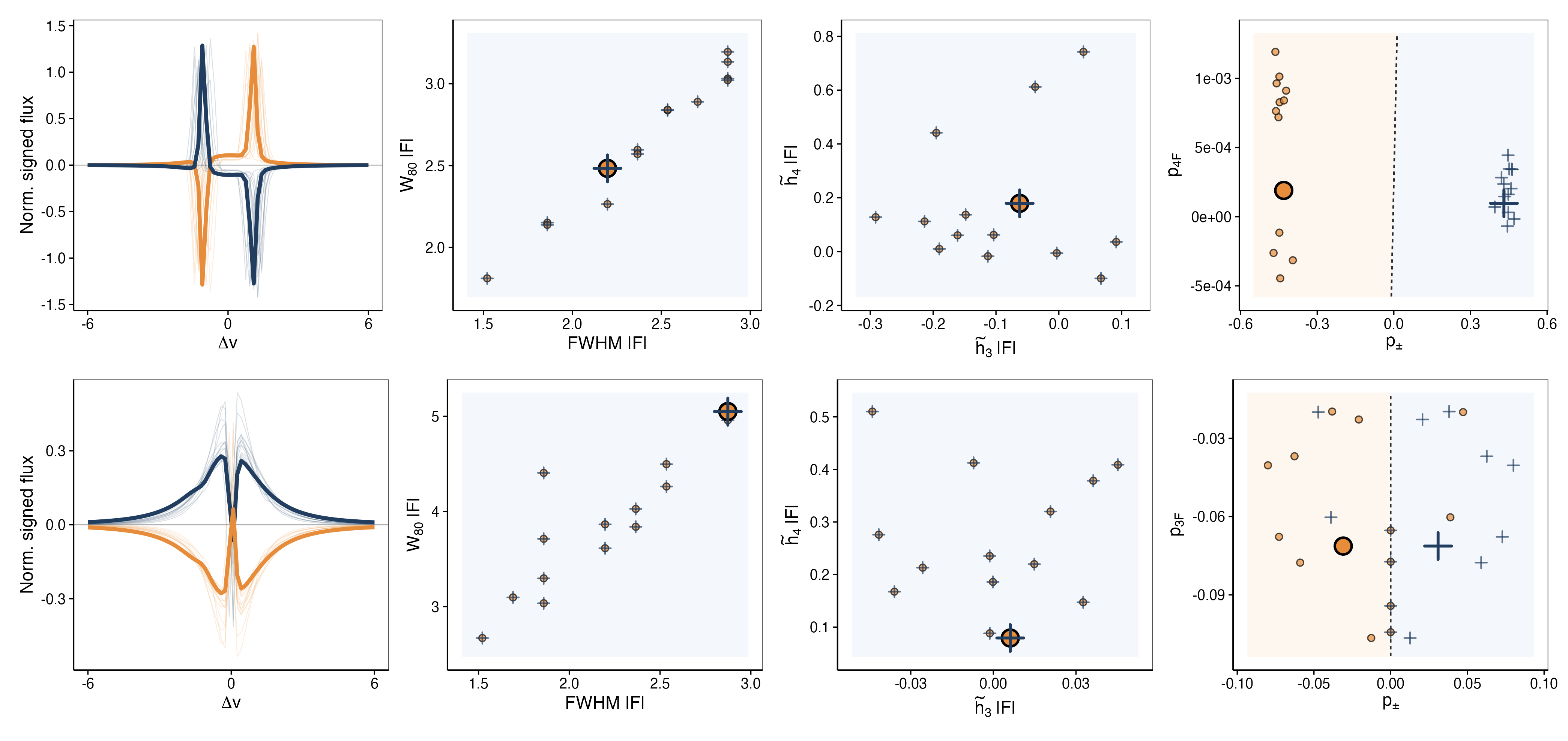}

\vspace{0.4cm}

\includegraphics[width=\linewidth]{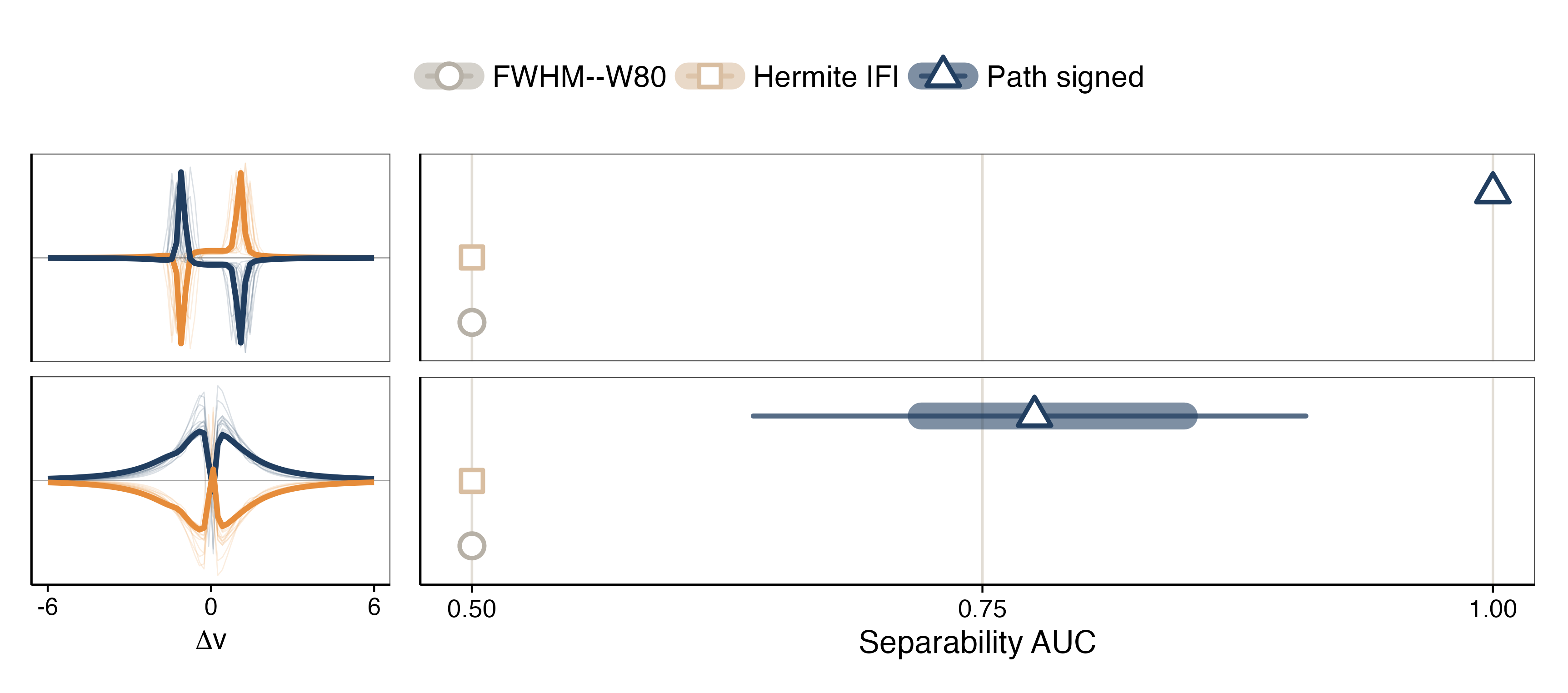}

\caption{Signed emission--absorption stress test for path-profile descriptors. The upper panel compares synthetic profile families with identical absolute-flux morphology but reversed emission--absorption ordering. Classical summaries are computed from absolute flux $(|F|)$, whereas the path plane uses the signed ordered curve. Since $|F|$-based summaries discard sign ordering, FWHM--$W_{80}$ and $\tilde h_3$--$\tilde h_4$ remain close to random separability. The signed path coordinates retain this ordering and separate the families, most clearly for the P-Cygni/inverse case. The lower panel shows bootstrap AUCs. 
}
\label{fig:signed_emabs_stress_test}
\end{figure*}

As a more targeted stress test, we consider paired continuum-subtracted
profiles in which emission and absorption components are arranged so that the
unsigned profile morphology is nearly unchanged, while the signed blue-to-red
ordering is reversed. This differs from the previous P-Cygni-like examples,
where emission--absorption structure also changes global profile shape and can
therefore remain visible to conventional summaries such as Hermite-like
coordinates. Here the experiment is designed to suppress that route: after an
unsigned preprocessing step such as taking \(|F|\), the two families become
nearly indistinguishable by construction.
This is a regime in which many conventional line-profile summaries require an
additional preprocessing convention. One may summarize \(|F|\), clip the
negative part, analyze emission and absorption separately, or fit an explicit
signed parametric model. These are all valid choices, but they define different
objects to be measured. Path coordinates can instead be evaluated directly on
the signed ordered curve
\(\gamma(\Delta v)=(\Delta v,F(\Delta v))\). Figure
\ref{fig:signed_emabs_stress_test} illustrates that the signed path
coordinates retain the emission--absorption ordering and therefore separate
the profile families, while summaries computed on \(|F|\) lose the
discriminating sign-order information.
This experiment should not be interpreted as showing that Hermite-based
summaries are generally inadequate. In the previous P-Cygni-like examples,
where the class difference also affects global skewness, width, or peakedness,
Hermite-like coordinates can remain competitive. The stress test instead
isolates a specific failure mode of preprocessing-dependent summaries: when
the relevant information is the ordered arrangement of signed flux, reducing
the profile to an unsigned distribution can erase the discriminating structure
that path coordinates preserve directly.

\subsection{Kinematics in MaNGA  Integral-field spectroscopy}
\label{sec:ifu}

\begin{figure*}
\centering
\includegraphics[width=\linewidth]{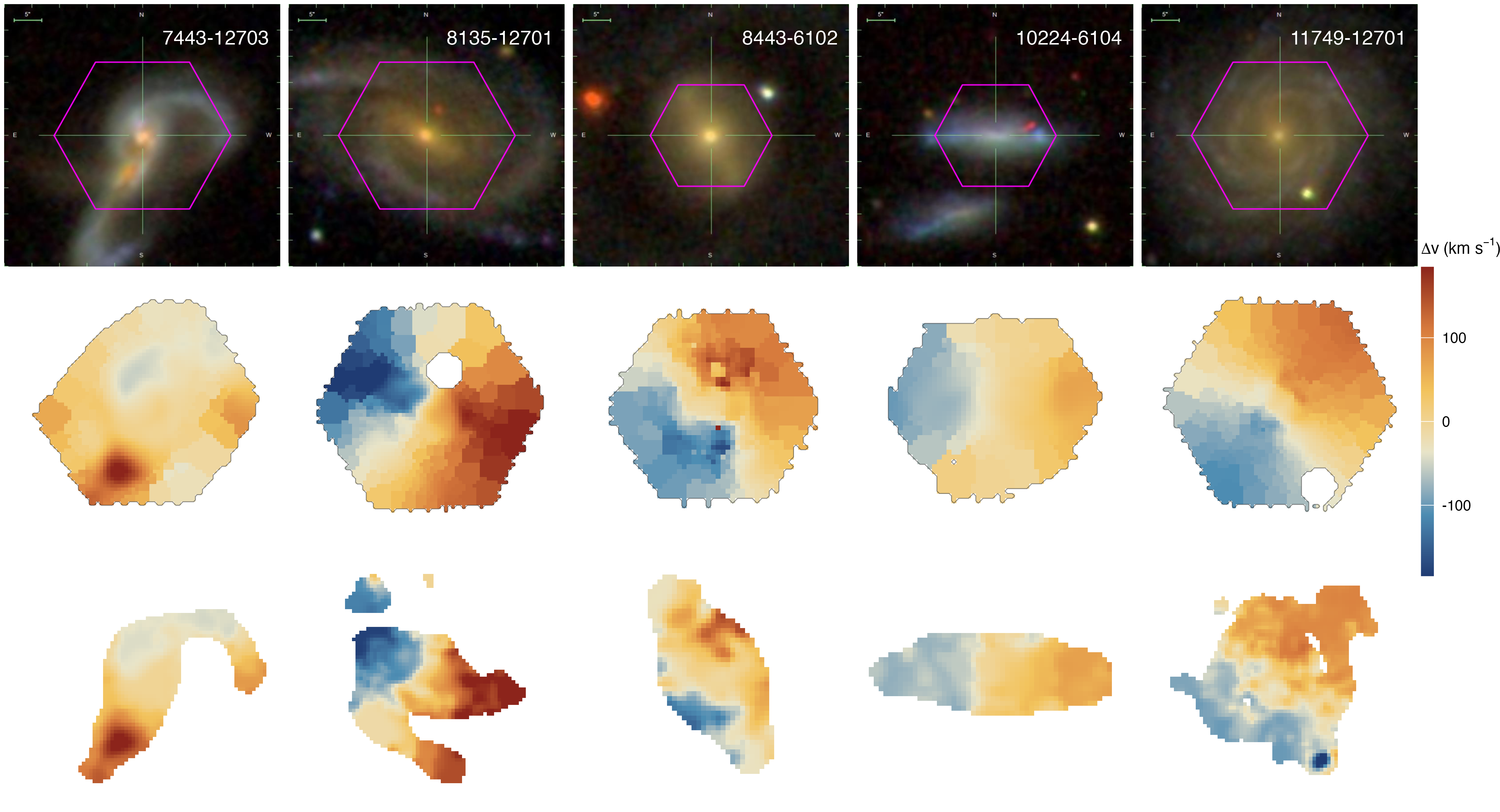}
\caption{Comparison between fiducial and path-segmented views of ionized-gas
kinematics in MaNGA galaxies. Top row: optical images with the MaNGA field of
view. Middle row: MaNGA DAP H\(\alpha\) velocity maps, shown as the
fiducial kinematic reference. Bottom row: velocities measured from spectra
stacked within regions selected by path-profile features
\((p_{3v},p_{3F},p_{4T})\). The bottom-row segmentation is not performed
directly on velocity. Instead, spaxels are grouped by ordered line-profile
morphology, and velocities are then measured from the stacked spectra in each
group. The comparison is intended as a qualitative proof of concept rather than as a quantitative validation of velocity recovery. Nevertheless, the path-selected regions recover coherent large-scale kinematic structure in a visually meaningful way, while retaining an interpretable morphology-based segmentation of the line profiles.
}
\label{fig:kinematics}
\end{figure*}

Integral-field spectroscopy provides a natural setting for path-based
line-profile diagnostics. Each spaxel contains a local spectrum, and each
emission line can be represented as an ordered curve in velocity--flux space.
After shifting to the galaxy rest frame and selecting a narrow window around a
given transition, we write the profile in spaxel \(i\) as $X_i(v)=\bigl(v,F_i(v)\bigr)$, 
where \(v\) is the velocity offset relative to the laboratory wavelength of
the transition, and \(F_i(v)\) is the local continuum-subtracted line profile.
This construction uses velocity as the coordinate system of the line profile,
not as a pre-measured kinematic observable. The spectra are first placed on a
common rest-frame velocity grid around a chosen transition, such as
H\(\alpha\) or [N\,{\sc ii}], using the systemic redshift of the galaxy. We do
not recenter each spaxel by its fitted line velocity before computing the path
features. Consequently, shifts of the local line profile relative to the
systemic line centre remain encoded in the curve \(F_i(v)\). The segmentation
therefore has access to kinematic information only through the ordered
morphology of the line profile, not through an externally supplied velocity
map.

We apply this construction to five morphologically diverse MaNGA galaxies
obtained through NED\footnote{\url{http://ned.ipac.caltech.edu/}} and
Marvin\footnote{\url{https://dr17.sdss.org/marvin/}}
\citep{Aguado2019ApJS}. The galaxies were selected to span interacting,
barred, spiral, and early-type systems with different ionized-gas morphologies and velocity fields. The corresponding MaNGA identifiers, common names, redshifts, and broad morphological classes are listed in
Table~\ref{tab:sample}.

For each spaxel, we compute the path coefficients introduced in
Table~\ref{tab:path_coefficients}. In the kinematic experiment, we use the
three-dimensional feature map
\begin{equation}
\phi_i \equiv \phi(X_i) =
\left(p_{3v,i},p_{3F,i},p_{4T,i}\right).
\end{equation}
The coefficient \(p_{3v}\) describes how the signed velocity--flux area of the
profile is distributed along the velocity coordinate, making it sensitive to
blue--red imbalance. The coefficient \(p_{3F}\) introduces a flux weighting,
helping to distinguish structure associated with the bright line core from
structure associated with fainter wings or shoulders. The fourth-order
coefficient \(p_{4T}\) captures higher-order bending of the velocity--flux
path, and is therefore sensitive to shoulders, double peaks, and
multi-component morphology.

Computed spaxel by spaxel, the components of \(\phi_i\) define maps of
line-profile morphology. These maps are analogous to eigenmaps only in the
limited sense that they display distinct channels of spectral variation across
the galaxy. Unlike principal components, however, their interpretation is fixed
before examining the data. A PCA component may mix line flux, centroid shifts,
width variations, and asymmetry depending on the cube and its preprocessing.
By contrast, the components of \(\phi_i\) retain the same geometric meaning
from object to object. They are therefore not variance-optimal components, but named diagnostics of ordered line shape.

\begin{table}
    \centering
    \caption{MaNGA galaxies used for the kinematic path-profile experiment.}
    \label{tab:sample}
    \begin{tabular}{llcc}
    \hline
    MaNGA ID & Object & $z$ & Morphology \\
    \hline
    7443-12703   & VV 705 / IRAS F15163+4255 & 0.0403 & Merger \\
    8135-12701   & UGC 3907                  & 0.0618 & SABc \\
    8443-6102    & UGC 8730                  & 0.0274 & (R)SB0 \\
    10224-6104   & MCG +00-07-007            & 0.0248 & Spiral \\
    11749-12701  & MCG +05-22-014            & 0.0417 & Sc \\
    \hline
    \end{tabular}
\end{table}

We then use \(\phi_i\) as the input space for a spectral-based segmentation via the {\sc capivara} package \citep{capivara2025,desouza2026} of the ionized-gas line profiles. In contrast to the original application of {\sc capivara}, where segmentation is
performed on the spectral cube or on broad spectral features, here the
clustering is restricted to the low-dimensional space of path-profile
descriptors. We therefore refer to this step as a kinematic {\sc capivara}
experiment: the algorithm is not given velocities directly, but groups spaxels according to the ordered morphology of their H\(\alpha\) profiles.
After clustering in \(\phi\)-space, the resulting labels are projected back
onto the MaNGA field of view. We then stack the H\(\alpha\) profiles within
each path-selected region and estimate a representative velocity from the
stacked spectrum. This produces a velocity map whose regions are selected by
line-profile morphology rather than by direct velocity fitting.

For each path-selected region \(g\), we construct a representative stacked
line profile by taking the median continuum-subtracted flux at each velocity
coordinate. We subtract a local baseline estimated from the edges of the line
window and measure a representative velocity from the first moment of the
positive, baseline-subtracted line profile,
\begin{equation}
v_g =
\frac{
\sum_k v_k F_g^+(v_k)
}{
\sum_k F_g^+(v_k)
},
\end{equation}
where \(F_g^+\) denotes the positive component of the baseline-subtracted
stacked profile. For visualization, the velocity maps show velocities after
subtracting the spatial median of the segment velocities. A single-Gaussian
line centre is also computed as a consistency check, but the maps shown here
use the non-parametric centroid because it does not assume a Gaussian line
shape.

Figure~\ref{fig:kinematics} summarizes this comparison. The top row shows the SDSS composite images and MaNGA fields of view for the five galaxies. The middle row shows the conventional H\(\alpha\) velocity field from the MaNGA DAP. The bottom row shows the region-wise velocity maps obtained after clustering spaxels in path-descriptor space, stacking the spectra within each path-selected region, and measuring a representative velocity from the stacked profile.

This experiment asks whether regions selected only from ordered line-profile morphology retain coherent kinematic information. The comparison is intended as a qualitative proof of concept rather than as a quantitative validation of velocity recovery. Nevertheless, the visual correspondence between the path-selected maps and the fiducial DAP velocity fields shows that the descriptors are not arbitrary shape summaries: they capture aspects of the line profiles that remain spatially and kinematically organized across the galaxy. The path-selected regions therefore trace coherent large-scale velocity structure while retaining an interpretable segmentation based on ordered line-profile morphology.

This provides a proof of concept for using path-based morphology as an intermediate representation for grouping spaxel-level spectra before subsequent kinematic or physical analysis. A detailed astrophysical interpretation of the recovered regions, including quantitative comparisons with fitted kinematics, non-Gaussian profiles, multi-component structure, and multi-line diagnostics, is deferred to future work.
Finally, path coefficients should be interpreted as summaries of preprocessed line morphology, not as noise-invariant observables. Small-scale fluctuations in the sampled profile can add artificial structure to the path. In practice, we therefore place profiles on a common velocity grid, normalize and lightly regularize them before comparison, and restrict attention to low-order terms, which are less sensitive to high-frequency fluctuations than higher-order signature coordinates.

\section{Conclusions}
\label{sec:conclusions}

We have introduced a geometric representation of astrophysical line profiles based on path signatures. In this framework, a continuum-subtracted spectral line is treated not only as a set of sampled flux values, but as an ordered curve traced across wavelength or velocity. This change of viewpoint makes the internal organization of the profile explicit and provides a compact set of interpretable geometric descriptors.

Using synthetic line profiles, we showed that these descriptors can distinguish morphologies that remain close in more compressed representations. The quantitative separability experiments indicate that path coordinates are particularly informative when the difference between profiles depends on the ordering and distribution of structure along the spectral axis. The method should therefore be understood as a morphology-aware representation rather than as a new physical classification scheme.

We then applied the same construction to MaNGA integral-field spectroscopy. For each spaxel, the H$\alpha$ profile was mapped into a low-dimensional path-descriptor space and clustered without using velocity or spatial position as direct inputs. After stacking the spectra within the resulting regions, the measured velocities recovered coherent large-scale kinematic structures broadly consistent with the MaNGA DAP fields. This result suggests that the path representation can serve as an intermediate layer between spaxel-level spectra and subsequent kinematic analysis, allowing spectra to be grouped according to their ordered structure before kinematic quantities are inferred.

Several limitations remain. The coefficients depend on the adopted line window, the quality of the continuum subtraction, the normalization, the sampling, and the noise properties of the data. Residual sky features, line blending, and uncertainties in the line centre can perturb the sampled path, particularly for weak lines and higher-order coordinates. Because path signatures are constructed from iterated integrals, a formal propagation of pixel-level variance arrays into non-linear descriptor covariance matrices is a mathematically distinct challenge. Similarly, the behaviour of these descriptors under systematic continuum-subtraction offsets and low signal-to-noise thresholds will require dedicated calibration. A natural next step is therefore a noise-aware formulation that propagates flux, continuum, and line-centre uncertainties into the path representation, allowing the resulting coefficients and segmentations to be interpreted probabilistically. For current applications, standard resampling techniques, such as bootstrapping or Monte Carlo flux perturbation, can already be used to estimate empirical uncertainties on the path descriptors.

In this introductory work, we have focused on establishing the geometric framework and demonstrating its baseline efficacy on synthetic profiles and high-quality MaNGA spectra. However, future incarnations of this framework won't be limited to line profiles. Many astronomical data products are naturally ordered, including light curves, transient spectra, radial profiles, spectral-energy distributions, polarization curves, and evolutionary tracks. Moreover, the framework need not remain restricted to two-dimensional paths. Higher-dimensional paths could follow several observables simultaneously, for example the joint evolution of flux and colour with time, or multiple emission lines across wavelength. In such settings, path signatures may encode not only how each observable changes, but also how their dependencies evolve along the ordered domain.

More broadly, this work points towards a form of interaction between mathematics and astronomical data analysis that goes beyond the conventional transfer of established mathematical tools into a new application domain. Modern astronomical data pose questions of dimensionality, heterogeneity, uncertainty, and intrinsic ordering, and these questions can themselves motivate new mathematical constructions, interpretations, and computational approximations. The resulting framework is therefore best viewed as a joint development: astronomical structure informs which aspects of path geometry are meaningful, while mathematics provides a language in which those structures can be made explicit.

In this sense, the path-signature view is less a replacement for existing methods than a change of perspective. It asks not only which values are present in the data, but how those values are ordered, connected, and jointly transformed. Treating astronomical data as paths therefore offers a way to move from describing isolated measurements towards describing the geometry of change itself.

\FloatBarrier
\section*{Acknowledgments}
\label{sec:acknowledgements}
SB would like to thank C.~Arias Abad, F.~Bischoff, J.~Faria Martins and D.~Lee for insightful conversations on signatures. RSS acknowledges support from the Conselho Nacional de Desenvolvimento Cientifico e Tecnologico (CNPq, Brazil, grants 446508/2024-1 and 315026/2025-1).

\section*{DATA AVAILABILITY}
This work uses data from the MaNGA DR17 data release, available at \href{https://www.sdss4.org/dr17/manga/}{https://www.sdss4.org/dr17/manga/}.

\bibliographystyle{mnras}
\bibliography{ref}

@ARTICLE{Herrera2020,
       author = {{Herrera-Camus}, R. and {Sturm}, E. and {Graci{\'a}-Carpio}, J. and {Veilleux}, S. and {Shimizu}, T. and {Lutz}, D. and {Stone}, M. and {Gonz{\'a}lez-Alfonso}, E. and {Davies}, R. and {Fischer}, J. and {Genzel}, R. and {Maiolino}, R. and {Sternberg}, A. and {Tacconi}, L. and {Verma}, A.},
        title = "{Molecular gas inflows and outflows in ultraluminous infrared galaxies at z {\ensuremath{\sim}} 0.2 and one QSO at z = 6.1}",
      journal = {\aap},
         year = 2020,
        month = jan,
       volume = {633},
          eid = {L4},
        pages = {L4},
          doi = {10.1051/0004-6361/201937109},
archivePrefix = {arXiv},
       eprint = {1912.05548},
 primaryClass = {astro-ph.GA},
       adsurl = {https://ui.adsabs.harvard.edu/abs/2020A&A...633L...4H}
}

@ARTICLE{Sakamoto2009,
       author = {{Sakamoto}, Kazushi and {Aalto}, Susanne and {Wilner}, David J. and {Black}, John H. and {Conway}, John E. and {Costagliola}, Francesco and {Peck}, Alison B. and {Spaans}, Marco and {Wang}, Junzhi and {Wiedner}, Martina C.},
        title = "{P Cygni Profiles of Molecular Lines Toward Arp 220 Nuclei}",
      journal = {\apjl},
         year = 2009,
        month = aug,
       volume = {700},
       number = {2},
        pages = {L104-L108},
          doi = {10.1088/0004-637X/700/2/L104},
archivePrefix = {arXiv},
       eprint = {0906.5197},
 primaryClass = {astro-ph.GA},
       adsurl = {https://ui.adsabs.harvard.edu/abs/2009ApJ...700L.104S}
}

@ARTICLE{Maschmann2023,
       author = {{Maschmann}, Daniel and {Halle}, Ana{\"e}lle and {Melchior}, Anne-Laure and {Combes}, Fran{\c{c}}oise and {Chilingarian}, Igor V.},
        title = "{The origin of double-peak emission-line galaxies: Rotating discs, bars, or galaxy mergers?}",
      journal = {\aap},
         year = 2023,
        month = feb,
       volume = {670},
          eid = {A46},
        pages = {A46},
          doi = {10.1051/0004-6361/202244746},
archivePrefix = {arXiv},
       eprint = {2212.02529},
 primaryClass = {astro-ph.GA},
       adsurl = {https://ui.adsabs.harvard.edu/abs/2023A&A...670A..46M}
}

@article{Nevin2018,
    author = {Nevin, R. and Comerford, J. M. and Müller-Sánchez, F. and Barrows, R. and Cooper, M. C.},
    title = {The origin of double-peaked narrow lines in active galactic nuclei – III. Feedback from biconical AGN outflows},
    journal = {Monthly Notices of the Royal Astronomical Society},
    volume = {473},
    number = {2},
    pages = {2160-2187},
    year = {2018},
    month = {01},
    issn = {0035-8711},
    doi = {10.1093/mnras/stx2433},
    url = {https://doi.org/10.1093/mnras/stx2433},
    eprint = {https://academic.oup.com/mnras/article-pdf/473/2/2160/21529865/stx2433.pdf},
}

@ARTICLE{Muller2015,
       author = {{M{\"u}ller-S{\'a}nchez}, F. and {Comerford}, J.~M. and {Nevin}, R. and {Barrows}, R.~S. and {Cooper}, M.~C. and {Greene}, J.~E.},
        title = "{The Origin of Double-peaked Narrow Lines in Active Galactic Nuclei. I. Very Large Array Detections of Dual AGNs and AGN Outflows}",
      journal = {\apj},
         year = 2015,
        month = nov,
       volume = {813},
       number = {2},
          eid = {103},
        pages = {103},
          doi = {10.1088/0004-637X/813/2/103},
archivePrefix = {arXiv},
       eprint = {1509.04291},
 primaryClass = {astro-ph.GA},
       adsurl = {https://ui.adsabs.harvard.edu/abs/2015ApJ...813..103M}
}

@ARTICLE{Mullaney2013,
       author = {{Mullaney}, J.~R. and {Alexander}, D.~M. and {Fine}, S. and {Goulding}, A.~D. and {Harrison}, C.~M. and {Hickox}, R.~C.},
        title = "{Narrow-line region gas kinematics of 24 264 optically selected AGN: the radio connection}",
      journal = {\mnras},
         year = 2013,
        month = jul,
       volume = {433},
       number = {1},
        pages = {622-638},
          doi = {10.1093/mnras/stt751},
archivePrefix = {arXiv},
       eprint = {1305.0263},
 primaryClass = {astro-ph.CO},
       adsurl = {https://ui.adsabs.harvard.edu/abs/2013MNRAS.433..622M}
}

@article{Veilleux05,
   author = "Veilleux, Sylvain and Cecil, Gerald and Bland-Hawthorn, Joss",
   title = "Galactic Winds", 
   journal= "Annual Review of Astronomy and Astrophysics",
   year = "2005",
   volume = "43",
   number = "Volume 43, 2005",
   pages = "769-826",
   doi = "https://doi.org/10.1146/annurev.astro.43.072103.150610",
   url = "https://www.annualreviews.org/content/journals/10.1146/annurev.astro.43.072103.150610",
   publisher = "Annual Reviews",
   issn = "1545-4282",
   type = "Journal Article",
  }

@ARTICLE{Heckman2000,
       author = {{Heckman}, Timothy M. and {Lehnert}, Matthew D. and {Strickland}, David K. and {Armus}, Lee},
        title = "{Absorption-Line Probes of Gas and Dust in Galactic Superwinds}",
      journal = {\apjs},
         year = 2000,
        month = aug,
       volume = {129},
       number = {2},
        pages = {493-516},
          doi = {10.1086/313421},
archivePrefix = {arXiv},
       eprint = {astro-ph/0002526},
 primaryClass = {astro-ph},
       adsurl = {https://ui.adsabs.harvard.edu/abs/2000ApJS..129..493H}
}

@inproceedings{Morley2024,
  author    = {{Morley}, Samuel},
  title     = {{RoughPy}: Streaming data is rarely smooth},
  title = {Proceedings of the Python in Science Conference},
  year      = {2024},
  doi       = {10.25080/DXWY3560}
}

@misc{esig2017,
  author       = {{Lyons}, Terry and {Maxwell}, David},
  title        = {{esig}: The esig Python package},
  year         = {2017},
  howpublished = {Python package documentation},
  url          = {https://esig.readthedocs.io/en/latest/},
  note         = {Python package for computing signatures and log-signatures of paths, powered by libalgebra}
}

@inproceedings{kidger2021signatory,
  title={{S}ignatory: differentiable computations of the signature and logsignature transforms, on both {CPU} and {GPU}},
  author={Kidger, Patrick and Lyons, Terry},
  booktitle={International Conference on Learning Representations},
  year={2021},
  note={\url{https://github.com/patrick-kidger/signatory}}
}

@article{Reizenstein2020,
author = {Reizenstein, Jeremy F. and Graham, Benjamin}, title = {Algorithm 1004: The Iisignature Library: Efficient Calculation of Iterated-Integral Signatures and Log Signatures}, 
year = {2020},
issue_date = {March 2020},
publisher = {Association for Computing Machinery}, 
address = {New York, NY, USA}, 
volume = {46}, 
number = {1}, 
issn = {0098-3500}, 
url = {https://doi.org/10.1145/3371237}, 
doi = {10.1145/3371237}, 
journal = {ACM Trans. Math. Softw.},
month = mar,
articleno = {8}, 
numpages = {21} 
}

@ARTICLE{desouza2026,
       author = {{de Souza}, Rafael S. and {Wille}, Andressa and {Shenoy}, Shravya and {Patil}, Aarya A. and {Krone-Martins}, Alberto and {Chies-Santos}, Ana L. and {Boehm}, Celine and {Rosa}, Reinaldo R. and {Pessi}, Thallis and {Ishida}, Emille E.~O. and {Dage}, Kristen C. and {Nakazono}, Lilianne and {Darc}, Phelipe and {Durgesh}, Rupesh},
        title = "{SAGUI: SED-based Segmentation of Multi-band Galaxy Images -- Application to JADES in GOODS-South}",
           journal = {\mnras},
         volume = {549},
    number = {4},
    pages = {1},
    year = {2026},
        month = jun,
          doi = {10.1093/mnras/stag1062},
archivePrefix = {arXiv},
       eprint = {2604.18812},
 primaryClass = {astro-ph.IM},
       adsurl = {https://ui.adsabs.harvard.edu/abs/2026MNRAS.tmp..999D}
}

@article{Refregier2003,
    author = {Refregier, Alexandre},
    title = {Shapelets — I. A method for image analysis},
    journal = {\mnras},
    volume = {338},
    number = {1},
    pages = {35-47},
    year = {2003},
    month = {01},
    issn = {0035-8711},
    doi = {10.1046/j.1365-8711.2003.05901.x},
    url = {https://doi.org/10.1046/j.1365-8711.2003.05901.x},
    eprint = {https://academic.oup.com/mnras/article-pdf/338/1/35/3832113/338-1-35.pdf},
}

@ARTICLE{Alsing2018,
       author = {{Alsing}, Justin and {Wandelt}, Benjamin},
        title = "{Generalized massive optimal data compression}",
      journal = {\mnras},
         year = 2018,
        month = may,
       volume = {476},
       number = {1},
        pages = {L60-L64},
          doi = {10.1093/mnrasl/sly029},
archivePrefix = {arXiv},
       eprint = {1712.00012},
 primaryClass = {astro-ph.CO},
       adsurl = {https://ui.adsabs.harvard.edu/abs/2018MNRAS.476L..60A}
}

@ARTICLE{Tegmark1997,
       author = {{Tegmark}, Max and {Taylor}, Andy N. and {Heavens}, Alan F.},
        title = "{Karhunen-Lo{\`e}ve Eigenvalue Problems in Cosmology: How Should We Tackle Large Data Sets?}",
      journal = {\apj},
         year = 1997,
        month = may,
       volume = {480},
       number = {1},
        pages = {22-35},
          doi = {10.1086/303939},
archivePrefix = {arXiv},
       eprint = {astro-ph/9603021},
 primaryClass = {astro-ph},
       adsurl = {https://ui.adsabs.harvard.edu/abs/1997ApJ...480...22T}
}

@article{Reichardt2001,
    author = {Reichardt, Christian and Jimenez, Raul and Heavens, Alan F.},
    title = {Recovering physical parameters from galaxy spectra using MOPED},
    journal = {\mnras},
    volume = {327},
    number = {3},
    pages = {849-867},
    year = {2001},
    month = {11},
    issn = {0035-8711},
    doi = {10.1046/j.1365-8711.2001.04768.x},
    url = {https://doi.org/10.1046/j.1365-8711.2001.04768.x},
    eprint = {https://academic.oup.com/mnras/article-pdf/327/3/849/2879592/327-3-849.pdf},
}

@ARTICLE{HeavensJimenezLahav2000,
       author = {{Heavens}, Alan F. and {Jimenez}, Raul and {Lahav}, Ofer},
        title = "{Massive lossless data compression and multiple parameter estimation from galaxy spectra}",
      journal = {\mnras},
         year = 2000,
        month = oct,
       volume = {317},
       number = {4},
        pages = {965-972},
          doi = {10.1046/j.1365-8711.2000.03692.x},
archivePrefix = {arXiv},
       eprint = {astro-ph/9911102},
 primaryClass = {astro-ph},
       adsurl = {https://ui.adsabs.harvard.edu/abs/2000MNRAS.317..965H}
}

@article{ChevyrevLyons2016,
  title   = {Characteristic functions of measures on geometric rough paths},
  author  = {Chevyrev, Ilya and Lyons, Terry},
  journal = {The Annals of Probability},
  volume  = {44},
  number  = {6},
  pages   = {4049--4082},
  year    = {2016}
}

@ARTICLE{capivara2025,
       author = {{de Souza}, Rafael S. and {Dahmer-Hahn}, Luis G. and {Shen}, Shiyin and {Chies-Santos}, Ana L. and {Chen}, Mi and {Rahna}, P.~T. and {Coelho}, Paula and {Riffel}, Rog{\'e}rio and {Ye}, Renhao and {Tahmasebzadeh}, Behzad},
        title = "{CAPIVARA: a spectral-based segmentation method for IFU data cubes}",
      journal = {\mnras},
         year = 2025,
        month = jun,
       volume = {539},
       number = {4},
        pages = {3166-3179},
          doi = {10.1093/mnras/staf688},
archivePrefix = {arXiv},
       eprint = {2410.21962},
 primaryClass = {astro-ph.GA},
       adsurl = {https://ui.adsabs.harvard.edu/abs/2025MNRAS.539.3166S}
}

@article{BOEDIHARDJO2016,
title = {The signature of a rough path: Uniqueness},
journal = {Advances in Mathematics},
volume = {293},
pages = {720-737},
year = {2016},
issn = {0001-8708},
doi = {https://doi.org/10.1016/j.aim.2016.02.011},
url = {https://www.sciencedirect.com/science/article/pii/S0001870816301104},
author = {Horatio Boedihardjo and Xi Geng and Terry Lyons and Danyu Yang}
}

@article{HamblyLyons2010,
  author  = {Hambly, Ben and Lyons, Terry},
  title   = {Uniqueness for the signature of a path of bounded variation and the reduced path group},
  journal = {Annals of Mathematics},
  volume  = {171},
  number  = {1},
  pages   = {109--167},
  year    = {2010}
}

@article{Chen1958,
  title={INTEGRATION OF PATHS—A FAITHFUL REPRE- SENTATION OF PATHS BY NONCOMMUTATIVE FORMAL POWER SERIES},
  author={Kuo-Tsai Chen},
  journal={Transactions of the American Mathematical Society},
  year={1958},
  volume={89},
  pages={395-407},
  url={https://api.semanticscholar.org/CorpusID:122551462}
}

@ARTICLE{Aguado2019ApJS,
       author = {{Aguado}, D.~S. and {Ahumada}, Romina and {Almeida}, Andr{\'e}s and {Anderson}, Scott F. and {Andrews}, Brett H. and {Anguiano}, Borja and {Aquino Ort{\'\i}z}, Erik and {Arag{\'o}n-Salamanca}, Alfonso and {Argudo-Fern{\'a}ndez}, Maria and {Aubert}, Marie and {Avila-Reese}, Vladimir and {Badenes}, Carles and {Barboza Rembold}, Sandro and {Barger}, Kat and {Barrera-Ballesteros}, Jorge and {Bates}, Dominic and {Bautista}, Julian and {Beaton}, Rachael L. and {Beers}, Timothy C. and {Belfiore}, Francesco and {Bernardi}, Mariangela and {Bershady}, Matthew and {Beutler}, Florian and {Bird}, Jonathan and {Bizyaev}, Dmitry and {Blanc}, Guillermo A. and {Blanton}, Michael R. and {Blomqvist}, Michael and {Bolton}, Adam S. and {Boquien}, M{\'e}d{\'e}ric and {Borissova}, Jura and {Bovy}, Jo and {Brandt}, William Nielsen and {Brinkmann}, Jonathan and {Brownstein}, Joel R. and {Bundy}, Kevin and {Burgasser}, Adam and {Byler}, Nell and {Cano Diaz}, Mariana and {Cappellari}, Michele and {Carrera}, Ricardo and {Cervantes Sodi}, Bernardo and {Chen}, Yanping and {Cherinka}, Brian and {Choi}, Peter Doohyun and {Chung}, Haeun and {Coffey}, Damien and {Comerford}, Julia M. and {Comparat}, Johan and {Covey}, Kevin and {da Silva Ilha}, Gabriele and {da Costa}, Luiz and {Dai}, Yu Sophia and {Damke}, Guillermo and {Darling}, Jeremy and {Davies}, Roger and {Dawson}, Kyle and {de Sainte Agathe}, Victoria and {Deconto Machado}, Alice and {Del Moro}, Agnese and {De Lee}, Nathan and {Diamond-Stanic}, Aleksandar M. and {Dom{\'\i}nguez S{\'a}nchez}, Helena and {Donor}, John and {Drory}, Niv and {du Mas des Bourboux}, H{\'e}lion and {Duckworth}, Chris and {Dwelly}, Tom and {Ebelke}, Garrett and {Emsellem}, Eric and {Escoffier}, Stephanie and {Fern{\'a}ndez-Trincado}, Jos{\'e} G. and {Feuillet}, Diane and {Fischer}, Johanna-Laina and {Fleming}, Scott W. and {Fraser-McKelvie}, Amelia and {Freischlad}, Gordon and {Frinchaboy}, Peter M. and {Fu}, Hai and {Galbany}, Llu{\'\i}s and {Garcia-Dias}, Rafael and {Garc{\'\i}a-Hern{\'a}ndez}, D.~A. and {Garma Oehmichen}, Luis Alberto and {Geimba Maia}, Marcio Antonio and {Gil-Mar{\'\i}n}, H{\'e}ctor and {Grabowski}, Kathleen and {Gu}, Meng and {Guo}, Hong and {Ha}, Jaewon and {Harrington}, Emily and {Hasselquist}, Sten and {Hayes}, Christian R. and {Hearty}, Fred and {Hernandez Toledo}, Hector and {Hicks}, Harry and {Hogg}, David W. and {Holley-Bockelmann}, Kelly and {Holtzman}, Jon A. and {Hsieh}, Bau-Ching and {Hunt}, Jason A.~S. and {Hwang}, Ho Seong and {Ibarra-Medel}, H{\'e}ctor J. and {Jimenez Angel}, Camilo Eduardo and {Johnson}, Jennifer and {Jones}, Amy and {J{\"o}nsson}, Henrik and {Kinemuchi}, Karen and {Kollmeier}, Juna and {Krawczyk}, Coleman and {Kreckel}, Kathryn and {Kruk}, Sandor and {Lacerna}, Ivan and {Lan}, Ting-Wen and {Lane}, Richard R. and {Law}, David R. and {Lee}, Young-Bae and {Li}, Cheng and {Lian}, Jianhui and {Lin}, Lihwai and {Lin}, Yen-Ting and {Lintott}, Chris and {Long}, Dan and {Longa-Pe{\~n}a}, Pen{\'e}lope and {Mackereth}, J. Ted and {de la Macorra}, Axel and {Majewski}, Steven R. and {Malanushenko}, Olena and {Manchado}, Arturo and {Maraston}, Claudia and {Mariappan}, Vivek and {Marinelli}, Mariarosa and {Marques-Chaves}, Rui and {Masseron}, Thomas and {Masters}, Karen L. and {McDermid}, Richard M. and {Medina Pe{\~n}a}, Nicol{\'a}s and {Meneses-Goytia}, Sofia and {Merloni}, Andrea and {Merrifield}, Michael and {Meszaros}, Szabolcs and {Minniti}, Dante and {Minsley}, Rebecca and {Muna}, Demitri and {Myers}, Adam D. and {Nair}, Preethi and {Correa do Nascimento}, Janaina and {Newman}, Jeffrey A. and {Nitschelm}, Christian and {Olmstead}, Matthew D. and {Oravetz}, Audrey and {Oravetz}, Daniel and {Ortega Minakata}, Ren{\'e} A. and {Pace}, Zach and {Padilla}, Nelson and {Palicio}, Pedro A. and {Pan}, Kaike and {Pan}, Hsi-An and {Parikh}, Taniya and {Parker}, III, James and {Peirani}, Sebastien and {Penny}, Samantha and {Percival}, Will J. and {Perez-Fournon}, Ismael and {Peterken}, Thomas and {Pinsonneault}, Marc H. and {Prakash}, Abhishek and {Raddick}, M. Jordan and {Raichoor}, Anand and {Riffel}, Rogemar A. and {Riffel}, Rog{\'e}rio and {Rix}, Hans-Walter and {Robin}, Annie C. and {Roman-Lopes}, Alexandre and {Rose}, Benjamin and {Ross}, Ashley J. and {Rossi}, Graziano and {Rowlands}, Kate and {Rubin}, Kate H.~R. and {S{\'a}nchez}, Sebasti{\'a}n F. and {S{\'a}nchez-Gallego}, Jos{\'e} R. and {Sayres}, Conor and {Schaefer}, Adam and {Schiavon}, Ricardo P. and {Schimoia}, Jaderson S. and {Schlafly}, Edward and {Schlegel}, David and {Schneider}, Donald P. and {Schultheis}, Mathias and {Seo}, Hee-Jong and {Shamsi}, Shoaib J. and {Shao}, Zhengyi and {Shen}, Shiyin and {Shetty}, Shravan and {Simonian}, Gregory and {Smethurst}, Rebecca J. and {Sobeck}, Jennifer and {Souter}, Barbara J. and {Spindler}, Ashley and {Stark}, David V. and {Stassun}, Keivan G.},
        title = "{The Fifteenth Data Release of the Sloan Digital Sky Surveys: First Release of MaNGA-derived Quantities, Data Visualization Tools, and Stellar Library}",
      journal = {\apjs},
         year = 2019,
        month = feb,
       volume = {240},
       number = {2},
          eid = {23},
        pages = {23},
          doi = {10.3847/1538-4365/aaf651},
archivePrefix = {arXiv},
       eprint = {1812.02759},
 primaryClass = {astro-ph.IM},
       adsurl = {https://ui.adsabs.harvard.edu/abs/2019ApJS..240...23A}
}

@article{Trager2000,
  author  = {Trager, S. C. and Faber, S. M. and Worthey, Guy and Gonz{\'a}lez, J. Jes{\'u}s},
  title   = {The Stellar Population Histories of Local Early-Type Galaxies. I. Population Parameters},
  journal = {AJ},
  year    = {2000},
  volume  = {119},
  pages   = {1645--1676},
  doi     = {10.1086/301299},
  eprint  = {astro-ph/0001072},
  archivePrefix = {arXiv}
}

@article{Worthey1994,
  author  = {Worthey, Guy and Faber, S. M. and Gonzalez, J. Jesus and Burstein, D.},
  title   = {Old Stellar Populations. V. Absorption Feature Indices for the Complete Lick/IDS Sample of Stars},
  journal = {ApJS},
  year    = {1994},
  volume  = {94},
  pages   = {687--722},
  doi     = {10.1086/192087}
}

@article{vanDerMarelFranx1993,
  author  = {{van der Marel}, R. P. and Franx, M.},
  title   = {A New Method for the Identification of Non-Gaussian Line Profiles in Elliptical Galaxies},
  journal = {ApJ},
  year    = {1993},
  volume  = {407},
  pages   = {525--539},
  doi     = {10.1086/172534}
}

@article{CappellariEmsellem2004,
  author  = {Cappellari, Michele and Emsellem, Eric},
  title   = {Parametric Recovery of Line-of-Sight Velocity Distributions from Absorption-Line Spectra of Galaxies via Penalized Likelihood},
  journal = {PASP},
  year    = {2004},
  volume  = {116},
  pages   = {138--147},
  doi     = {10.1086/381875}
}

@article{Bu2015,
  author  = {Bu, Yude and Zhao, Gang and Luo, A.-Li and Pan, Jingchang and Chen, Yuqin},
  title   = {Restricted Boltzmann Machine: A Non-linear Substitute for PCA in Spectral Processing},
  journal = {A\&A},
  year    = {2015},
  volume  = {576},
  pages   = {A96},
  doi     = {10.1051/0004-6361/201424194}
}

@article{Wang2017,
  author  = {Wang, Ke and Guo, Ping and Luo, A.-Li},
  title   = {A New Automated Spectral Feature Extraction Method and Its Application in Spectral Classification and Defective Spectra Recovery},
  journal = {MNRAS},
  year    = {2017},
  volume  = {465},
  pages   = {4311--4324},
  doi     = {10.1093/mnras/stw2894}
}

@article{Sanchez2022,
  author  = {S{\'a}nchez, S. F. and Barrera-Ballesteros, J. K. and Lacerda, E. and Mej{\'i}a-Narvaez, A. and Camps-Fari{\~n}a, A. and Bruzual, G. and Espinosa-Ponce, C. and Rodr{\'i}guez-Puebla, A. and Calette, A. R. and Ibarra-Medel, H. and Avila-Reese, V. and Hernandez-Toledo, H. and Bershady, M. A. and Cano-Diaz, M. and Munguia-Cordova, A. M.},
  title   = {SDSS-IV MaNGA: pyPipe3D Analysis Release for 10,000 Galaxies},
  journal = {ApJS},
  year    = {2022},
  volume  = {262},
  number  = {2},
  pages   = {36},
  doi     = {10.3847/1538-4365/ac7b8f}
}

@article{ConnollySzalay1999,
  author  = {Connolly, A. J. and Szalay, A. S.},
  title   = {A Robust Classification of Galaxy Spectra: Dealing with Noisy and Incomplete Data},
  journal = {AJ},
  year    = {1999},
  volume  = {117},
  pages   = {2052--2062},
  doi     = {10.1086/300839}
}

@article{Yip2004a,
  author  = {Yip, C. W. and Connolly, A. J. and Szalay, A. S. and Bud{\'a}vari, T. and SubbaRao, M. and Frieman, J. A. and Nichol, R. C. and Hopkins, A. M. and York, D. G. and Okamura, S. and Brinkmann, J. and Csabai, I. and Thakar, A. R. and Fukugita, M. and Ivezi{\'c}, {\v Z}.},
  title   = {Distributions of Galaxy Spectral Types in the Sloan Digital Sky Survey},
  journal = {AJ},
  year    = {2004},
  volume  = {128},
  pages   = {585--609},
  doi     = {10.1086/422429}
}

@article{Yip2004b,
  author  = {Yip, C. W. and Connolly, A. J. and Vanden Berk, D. E. and Ma, Z. and Frieman, J. A. and SubbaRao, M. and Szalay, A. S. and Richards, G. T. and Hall, P. B. and Schneider, D. P. and Hopkins, A. M. and Trump, J. and Brinkmann, J.},
  title   = {Spectral Classification of Quasars in the Sloan Digital Sky Survey: Eigenspectra, Redshift, and Luminosity Effects},
  journal = {AJ},
  year    = {2004},
  volume  = {128},
  pages   = {2603--2630},
  doi     = {10.1086/425626}
}

@ARTICLE{Connolly1995,
       author = {{Connolly}, A.~J. and {Szalay}, A.~S. and {Bershady}, M.~A. and {Kinney}, A.~L. and {Calzetti}, D.},
        title = "{Spectral Classification of Galaxies: an Orthogonal Approach}",
      journal = {\aj},
         year = 1995,
        month = sep,
       volume = {110},
        pages = {1071},
          doi = {10.1086/117587},
archivePrefix = {arXiv},
       eprint = {astro-ph/9411044},
 primaryClass = {astro-ph},
       adsurl = {https://ui.adsabs.harvard.edu/abs/1995AJ....110.1071C}
}

@article{Bayes1763,
    author = {Bayes, Thomas},
    title = {LII. An essay towards solving a problem in the doctrine of chances. By the late Rev. Mr. Bayes, F. R. S. communicated by Mr. Price, in a letter to John Canton, A. M. F. R. S},
    journal = {Philosophical Transactions},
    number = {53},
    pages = {370-418},
    year = {1763},
    month = {12},
    issn = {0260-7085},
    doi = {10.1098/rstl.1763.0053},
    url = {https://doi.org/10.1098/rstl.1763.0053},
    eprint = {https://royalsocietypublishing.org/rstl/article-pdf/doi/10.1098/rstl.1763.0053/1458453/rstl.1763.0053.pdf},
}

@book{grattan1997fontana,
  title={The Fontana History of the Mathematical Sciences: The Rainbow of Mathematics},
  author={Grattan-Guinness, I.},
  isbn={9780006861799},
  lccn={97215849},
  series={Fontana history of science},
  url={https://books.google.com.br/books?id=jmRvQgAACAAJ},
  year={1997},
  publisher={Fontana Press}
}

@book{Lagrange1788,
  author    = {Lagrange, Joseph-Louis},
  title     = {M{\'e}canique analytique},
  publisher = {Chez la veuve Desaint},
  address   = {Paris},
  year      = {1788}
}

@article{Hamilton1834,
  author  = {Hamilton, William Rowan},
  title   = {On a General Method in Dynamics; by which the Study of the Motions of All Free Systems of Attracting or Repelling Points is Reduced to the Search and Differentiation of One Central Relation, or Characteristic Function},
  journal = {Philosophical Transactions of the Royal Society of London},
  volume  = {124},
  pages   = {247--308},
  year    = {1834},
  doi     = {10.1098/rstl.1834.0017}
}

@article{Einstein1916,
  author  = {Einstein, Albert},
  title   = {Die Grundlage der allgemeinen Relativit\"atstheorie},
  journal = {Annalen der Physik},
  volume  = {354},
  number  = {7},
  pages   = {769--822},
  year    = {1916},
  doi     = {10.1002/andp.19163540702}
}

@article{shannon48,
  author = {Shannon, Claude Elwood},
  journal = {The Bell System Technical Journal},
  pages = {379--423},
  title = {A Mathematical Theory of Communication},
  url = {http://plan9.bell-labs.com/cm/ms/what/shannonday/shannon1948.pdf},
  urldate = {2003-04-22},
  volume = 27,
  year = 1948
}

@book{starck2006,
  author    = {Starck, Jean-Luc and Murtagh, Fionn},
  title     = {Astronomical Image and Data Analysis},
  year      = {2006},
  publisher = {Springer},
  address   = {Berlin, Heidelberg},
  edition   = {2}
}

@book{laplace1820,
  title={Th{\'e}orie analytique des probabilit{\'e}s},
  author={de Laplace, P.S.},
  series={Oeuvres compl{\`e}tes de Laplace},
  year={1820},
  publisher={Courcier}
}

@book{fourier1822,
  title={Th{\'e}orie analytique de la chaleur},
  author={Fourier, J.B.J.},
  series={Landmarks of Science},
  year={1822},
  publisher={Didot}
}

@article{Lyons1998,
author = {Lyons, Terry J.},
journal = {Revista Matemática Iberoamericana},
language = {eng},
number = {2},
pages = {215-310},
title = {Differential equations driven by rough signals.},
url = {http://eudml.org/doc/39555},
volume = {14},
year = {1998},
}

@article{CidFernandes2005,
   author = {{Cid Fernandes}, R. and {Mateus}, A. and {Sodr{\'e}}, L. and 
	{Stasi{\'n}ska}, G. and {Gomes}, J.~M.},
    title = "{Semi-empirical analysis of Sloan Digital Sky Survey galaxies - I. Spectral synthesis method}",
  journal = {\mnras},
   eprint = {astro-ph/0412481},
     year = 2005,
    month = apr,
   volume = 358,
    pages = {363-378},
      doi = {10.1111/j.1365-2966.2005.08752.x},
   adsurl = {http://adsabs.harvard.edu/abs/2005MNRAS.358..363C}
}

@article{lee2023random,
    title={Random surfaces and higher algebra},
    author={Lee, Darrick and Oberhauser, Harald},
    archivePrefix = {arXiv},
    eprint = {2311.08366},
    journal = "",
    primaryClass = {math.PR},
    year={2023}
}

@article{lee2024surface,
    title={The surface signature and rough surfaces},
    author={Lee, Darrick},
    year={2024},
    journal = "",
    archivePrefix = {arXiv},
    eprint = {2406.16857},
    primaryClass = {math.FA},
}

\appendix

\section{The path signature as a universal connection}
\label{app:gauge}
This appendix gives additional geometric background for the path signature.
It is intended to illuminate the various properties of the path signature, including its invariance under reparameterization, translation and thin homotopy, its multiplicativity and its ability to distinguish paths in $\bbR^n$ (up to reparameterization, translation and thin homotopy).

In particular, we advocate the perspective of viewing the path signature as a parallel transport construction: it is the universal translation-invariant parallel transport on $\bbR^n$.
This appendix aims to illustrate the meaning of this statement and aid the reader in further understanding the path signature.
As such, it emphasises exposition and concepts over mathematical rigour.

\subsection{The parallel transport of a connection}
\label{sec:PT of connection}

Connections are mathematical structures defined on `principal bundles'.
The Levi-Civita connection in general relativity, and gauge fields in particle physics are prominent examples of where such connections play important roles in physical contexts.
Here we will only import a minimal part of this theory, tailored to our purposes.
In particular, we will be working only on the manifold $\bbR^n$ as the base space.
In particular, every connection on $\bbR^n$ lives on a manifold $P$ which comes with a smooth map $\pi \colon P \to \bbR^n$, and whose fibres are (non-canonically) isomorphic to the structure group $G$ of the connection.
This $G$ is a Lie group, and we denote its Lie algebra by $\mathfrak{g}$.
The full mathematical structure of the map $P \to M$ is that of a principal $G$-bundle, but we will not need to fully unravel this term here.

It turns out that, since we are working on $M = \bbR^n$, every principal bundle $P \to \bbR^n$ admits a global section%
\footnote{
    In general, on manifolds $M$ of non-trivial topology, sections of principal bundles $\pi \colon P \to M$ exist only locally, and to obtain a global description of connections on $P$ one has to patch together many of these local pictures.
},
i.e.~a smooth map $s \colon \bbR^n \to P$ such that the composite map $\pi \circ s$ is the identity map on $\bbR^n$.
Each such choice of a section of $P$ translates any connection on $P$ into a differential $1$-form $A = A_\mu dx^\mu$ on $\bbR^n$ with values in the Lie algebra $\mathfrak{g}$.
In fact, this induces a 1-to-1 correspondence between connections on the bundle $P$ and differential 1-forms on $\bbR^n$ with values in $\frg$.
However, this identification is not canonical, but depends on the choice of global section: if we keep the connection on $P$ fixed, but pass to a different global section $s'$ of $P$, then the resulting $1$-form $A'$ which represents the connection will differ from $A$ precisely by a gauge transformation.

Every connection on a principal bundle $\pi \colon P \to \bbR^n$ further corresponds to a `parallel transport' on $P$.
We briefly recall this notion and how it is induced by a connection.
If we have a smooth%
\footnote{
    The smoothness condition can be relaxed so that we require the path to only be piecewise smooth.
    For the signature, one can even work with much less regular paths called `rough' paths~\cite{Lyons1998}.
}
path $\gamma \colon [0,1] \to \bbR^n$ together with a point $p \in P$ in the fibre over the initial point $\gamma(0)$, the parallel transport associated to the connection defines a unique lift of $\gamma$ to a smooth path $\hat{\gamma} \colon [0,1] \to P$ which starts at $p$.
The path $\hat{\gamma}$ is the `horizontal lift', or `covariantly constant' lift of $\gamma$ to $P$.
For later use, we record that if $\gamma$ is a closed loop in $M$, then $\hat{\gamma}(1)$ and $\hat{\gamma}(0) = p$ lie in the same fibre of $P$ and so---by invoking the mathematical properties of principal bundles, which we have black-boxed here---there is a unique element $g \in G$ of the structure group such that
\begin{equation}
    \hat{\gamma}(1) = \hat{\gamma}(0) \cdot g\,.
\end{equation}
We call this element $g \in G$ the `holonomy' of the connection along $\gamma$ based at $p$.

\begin{figure}
    \centering
    \includegraphics[width=0.95 \linewidth]{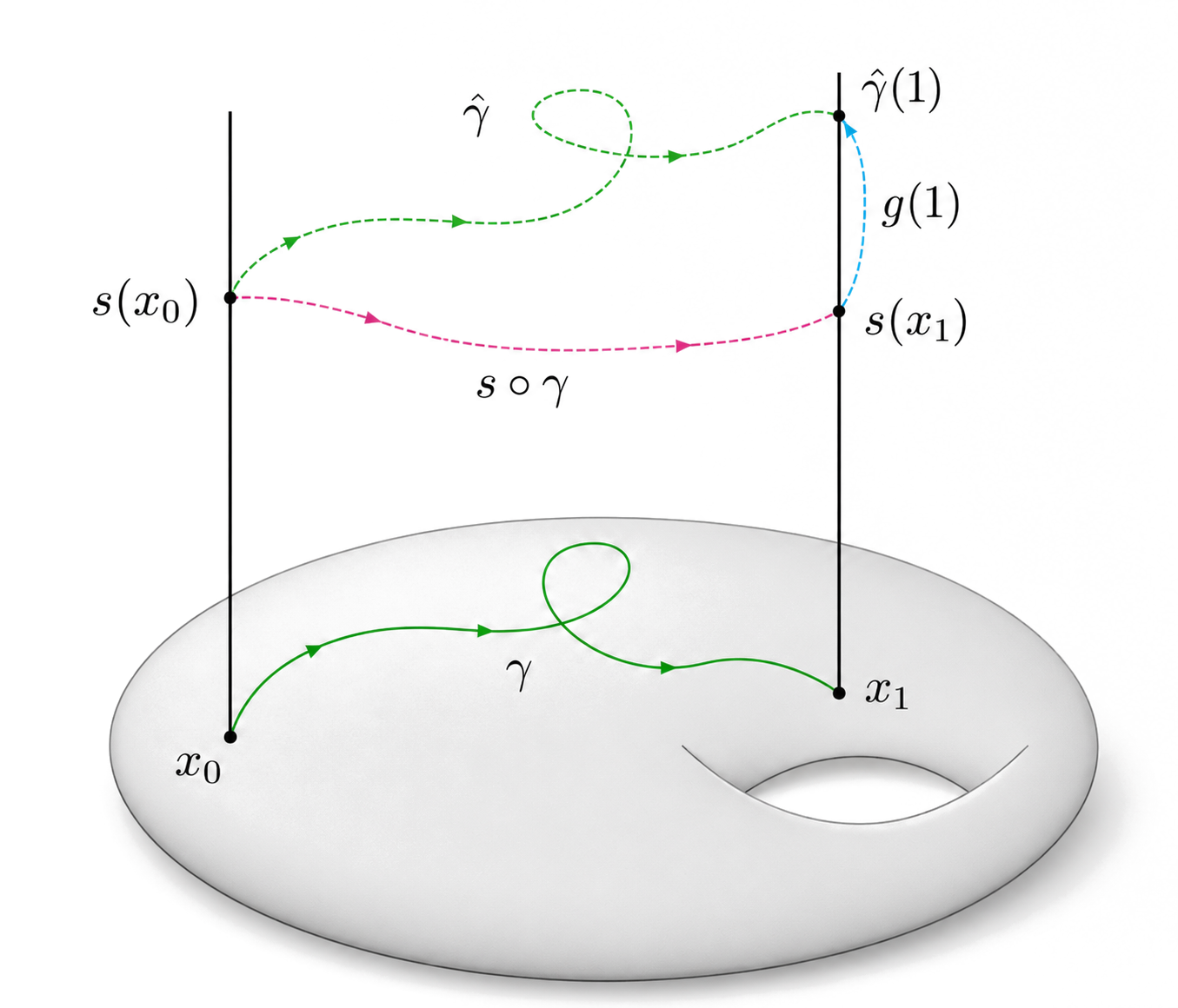}
   \caption{Illustration of parallel transport along a path $\gamma$ in the base manifold together with a chosen global section $s$. The path $\gamma$ runs from $x_0$ to $x_1$, while the section selects points $s(x_0)$ and $s(x_1)$ in the corresponding fibres. Starting from $s(x_0)$, parallel transport along $\gamma$ defines a horizontal lift $\hat{\gamma}$ in the total space. Relative to the point $s(\gamma(t))$, i.e.~using the section to lift the path $\gamma$, the actual parallel transport differs from $s(\gamma(t))$ by the action of a unique group element $g(t)$, so that $\hat{\gamma}(t) = s(\gamma((t)) \cdot g(t)$. Thus, parallel transport along $\gamma$ is represented, with respect to the section $s$, by a path in the structure group.}
    \label{fig: PT and section}
\end{figure}

Suppose now that we have chosen and fixed a section $s \colon \bbR^n \to P$, so that we can work purely in terms of the representation of the connection as a $1$-form $A$ valued in the Lie algebra of $G$.
Then, given a path $\gamma \colon [0,1] \to \bbR^n$ as above, we may take $p = s(\gamma(0))$ and try to compute the lift $\hat{\gamma}$ in terms of $A$.
By the properties of principal bundles it holds true that, for every time $t \in [0,1]$, the lift $\hat{\gamma}$ and the section $s$ are now related:
there is a unique smooth path $g \colon [0,1] \to G$ such that (see Figure~\ref{fig: PT and section})
\begin{equation}
\label{eq:group-valued PT via section}
    \hat{\gamma}(t) = s \big( \gamma(t) \big) \cdot g(t)
    \quad
    \forall\, t \in [0,1]\,,
\end{equation}
where we have used that the structure group $G$ acts on each fibre of $P$ from the right.
That is, to compute the lift $\hat{\gamma}$, we can equivalently compute the path $g \colon [0,1] \to G$.

We now assume, for the sake of exposition, that $G$ is a matrix Lie group (such as $GL(n)$, $SO(n)$, $SU(n)$, etc.).
Then, a path $g \colon [0,1] \to G$ satisfies Equation~\eqref{eq:group-valued PT via section}, and hence captures the parallel transport along $\gamma$, if and only if it satisfies the differential equation
\begin{equation}
\label{eq:PT ODE for G}
    \dot{g}(t) = A_{|\gamma(t)} \big( \dot{\gamma}(t) \big) \cdot g(t)
    = \dot{\gamma}^\mu(t)\, A_\mu(\gamma(t))\, g(t)\,,
\end{equation}
where both sides of the equation live in $G$ (for each $t \in [0,1]$).
Here, the differential $1$-form $A$ is evaluated at the point $\gamma(t) \in \bbR^n$ and then applied to the tangent vector $\dot{\gamma}(t)$ to $\gamma$ at time $t$.
Since later on we will only be interested in the case where $A$ is constant on $\bbR^n$, as well as to simplify our notation, we will drop the dependency of $A_{|x}$ on the point $x \in \bbR^n$ from the notation from now on.

Let us provide a heuristic description for how one can solve the differential equation~\eqref{eq:PT ODE for G}.
For very small time steps $\Delta t$, we can write
\begin{equation}
\label{eq:PT and connection}
    \dot{g}(0) = \frac{g(\Delta t) - g(0)}{\Delta t}
    = A \big( \dot{\gamma}(0) \big)\, g(0)\,.
\end{equation}
Solving for $g(\Delta t)$, we obtain
\begin{equation}
\label{eq:g(Delta t)---first step}
    g(\Delta t) = g(0) + A \big( \dot{\gamma}(0) \big)\, g(0) \cdot \Delta t\,.
\end{equation}
Using this, we can compute the value $g(2 \Delta t)$, after the second time step, as
\begin{align}
    g(2 \Delta t)
    &= g(\Delta t) + A \big( \dot{\gamma}(\Delta t) \big) \cdot g(\Delta t) \cdot \Delta t
    \\
    &= g(0) + A \big( \dot{\gamma}(0) \big) \cdot g(0) \cdot \Delta t
    \notag
    \\
    &\quad + A \big( \dot{\gamma}(\Delta t) \big)
    \cdot \Big( g(0) + A \big( \dot{\gamma}(0) \big)\, g(0) \cdot \Delta t \Big) \cdot \Delta t
    \notag
    \\
    &= g(0) + \Big( A \big( \dot{\gamma}(\Delta t) \big) + A \big( \dot{\gamma}(0) \big) \Big)\, g(0) \cdot \Delta t
    \notag
    \\
    &\quad + A \big( \dot{\gamma}(\Delta t) \big) \cdot A \big( \dot{\gamma}(0) \big) \cdot g(0) \cdot \Delta t^2\,.
    \notag
\end{align}
Further iterating this process and taking the limit $\Delta t \mapsto 0$ of infinitely many and infinitesimally small time steps, terms as in the second summand in the last line turn into a Riemann integral, whereas terms like the final summand above approximate products of Riemann integrals.
Note, however, that in the final term above---and all other terms containing products of the form $A(\dot{\gamma}(n \cdot \Delta t))\, A(\dot{\gamma}((n-1) \cdot \Delta t)$ that appear in further iterations---the product is always `time-ordered', i.e.~instances of $A$ evaluated on later tangent vectors always appear on the left.
In the limit, we thus obtain the expression
\begin{align}
	\label{eq: iterated integral formula for Pexp}
    g(t) &= g(0)
    \\
    & + \sum_{n = 1}^\infty \int_{t_n \geq \cdots \geq t_1} A \big( \dot{\gamma}(t_n) \big) \cdots A \big( \dot{\gamma}(t_1) \big) \cdot g(0)\, dt_1 \cdots dt_n\,.
    \notag
\end{align}
In particular, choosing $s(\gamma(0))$ as the start point for the lift $\hat{\gamma}$, we can always take $g(0)$ to be the unit element.
The resulting expression is also called the `path-ordered exponential' of $A(\dot{\gamma})$ and written as
\begin{equation}
	\label{eq: Pexp formula for PT}
    g(t) = P\exp \left( \int_t A \big( \dot{\gamma}(t) \big)\, dt \right)\,.
\end{equation}
It provides a closed formula for the parallel transport of the connection along any smooth path $\gamma$ (relative to the section $s$).

In particular, a parallel transport on $\bbR^n$ induces a map which assigns to each smooth path $\gamma$ in $\bbR^n$ a group element $g(1) \in G$.
The link to the path signature is now to view such an assignment as a potential feature map for path-like data.
It leads us to the question what part of the information contained in the path $\gamma$ the parallel transport along $\gamma$ remembers and thus whether we can build a parallel transport which is optimal in the sense that it remembers everything about the path, i.e.~which is able to distinguish paths in $\bbR^n$.

\subsection{The path signature as a parallel transport}

The path signature can be interpreted as a parallel transport of a connection on $\bbR^n$ that is `universal' in a precise sense (see, for instance,~\cite{ChevyrevLyons2016, lee2023random, lee2024surface}).
In~\cite{lee2024surface} the relation was written as a direct, algebraic comparison of the path signature and the path-ordered exponential describing the parallel transport of an arbitrary translation-invariant connection on $\bbR^n$.
Here we instead derive this relationship from a geometric perspective, in the hope to clarify which bundle and structure group underlie the signature.
In particular, this perspective readily explains the key result of \citeauthor{HamblyLyons2010}---at least in the case of piecewise smooth paths---that the path signature can distinguish paths up to thin homotopy or `tree-like equivalence' (see below for more details).
While our discussion is not at the level of rigour required of a mathematical proof, it provides a new perspective and understanding of the path signature which sheds valuable light on the global geometric nature of the signature and its properties.

Let us take the following question as the motivating problem for the discussion in this section:

\begin{quote}
  \emph{Can we build a parallel transport on $\bbR^n$ that can distinguish any two smooth paths?}
\end{quote}

\begin{figure}
    \centering
    \includegraphics[width=0.75 \linewidth]{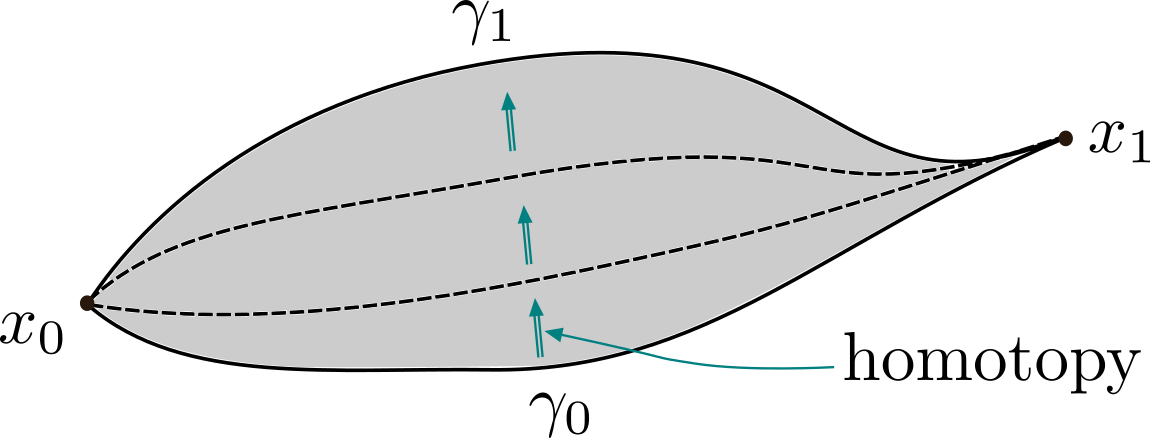}
    \caption{
        A homotopy (i.e.~a smooth deformation) between two paths $\gamma_0$ and $\gamma_1$ with fixed endpoints.
        In the process of deforming the path $\gamma_0$ into $\gamma_1$, the homotopy sweeps out a surface (grey shaded area).
        The homotopy is `thin' if this surface is degenerate, i.e.~only forms a 1-dimensional object, or has zero surface area.
    }
    \label{fig: homotopy of paths}
\end{figure}

It is known that the parallel transport of any connection along two paths $\gamma_0$, $\gamma_1$ with the same start points and the same end points will agree whenever the paths are `thinly homotopic' to each other or one is a reparameterization of the other.
This is the case whenever there exists a smooth deformation of $\gamma_0$ into $\gamma_1$ with two properties: it does not alter the start or end points of the paths, and the surface it swipes out in the process is degenerate, i.e.~has zero surface area%
\footnote{
    More precisely, the deformation is a map $h \colon [0,1]^2 \to M$ such that $h(0,-) = \gamma_0$ and $h(1,-) = \gamma_1$, as well as $h(-,0)$ and $h(-,1)$ are constant paths ($h$ leaves the start and end points fixed).
    The degeneracy of the surface amounts to the property that the vectors $\frac{\partial}{\partial s} h(s,t)$ and $\frac{\partial}{\partial t} h(s,t)$ are linearly dependent at each point $(s,t) \in [0,1]$.
}
(see Figure~\ref{fig: homotopy of paths} for an illustration).

Consequently, to have any chance for a positive answer, we need to refine our initial question as follows:

\begin{quote}
\emph{Can we build a parallel transport on $\bbR^n$ that distinguishes any two smooth paths up to thin homotopy and reparameterization?}
\end{quote}

In fact, the method we will produce below (which is the path signature) will further identify paths in $\bbR^n$ that differ only by a translation.
We will address this question at two levels.
First, we will discuss the actual parallel transport itself, including the bundle it is defined on, and second we will extract the associated connection form.
The associated expression for the parallel transport via the path-ordered exponential will then reproduce the path signature.

For the first step, we begin by identifying the bundle $P \to \bbR^n$ on which our hypothetical parallel transport is defined.
Consider the bundle
\begin{equation}
    \ev_1 \colon P' \bbR^n \longrightarrow \bbR^n
\end{equation}
defined as follows.
The space $P' \bbR^n$
is the space of all smooth paths in $\bbR^n$, modulo the equivalence relation of thin homotopy with fixed endpoints.
The map $\ev_1$ acts on an equivalence class $[\beta]$ as $[\beta] \mapsto \beta(1)$, i.e.~it evaluates the path $\beta$ at time $t = 1$%
\footnote{
    The space $P' \bbR^n$ can be given a natural smooth structure, for instance by viewing it as a diffeological space.
    Then, the endpoint evaluation map $\ev_1$ becomes a smooth map.
}.

The structure group of this bundle is the group of thin based paths $P'_0 \bbR^n$, whose elements are the thin-homotopy classes of paths in $\bbR^n$ which end at the origin, i.e.~classes $[\alpha]$ of smooth paths such that $\alpha(1) = 0$.
The multiplication of two elements in this group is defined as
\begin{equation}
    \alpha \cdot \alpha' = \alpha * (\alpha' + \alpha(0))\,,
\end{equation}
where the star denotes concatenation of paths (cf.~Equation~\eqref{eq: path concatenation}), and $\alpha + \alpha(0)$ is the path $\alpha$ shifted by the vector $\alpha(0)$ (the shift ensures that concatenation is defined).

The group $P'_0 \bbR^n$ acts on the space $P'\bbR^n$ via concatenation (after an appropriate translation), via
\begin{align}
    P'\bbR^n \times P'_0\bbR^n &\longrightarrow P'\bbR^n\,,
    \\
    (\beta, \alpha) &\longmapsto \beta \triangleleft \alpha = \beta * (\alpha + \beta(0))\,.
    \notag
\end{align}
Note that, for each pair $(\beta, \beta')$ of paths ending at the same vector $x \in \bbR^n$, we have that
\begin{equation}
    \beta = \beta' \triangleleft (\beta'^{-1} * \beta - \beta'(0))
\end{equation}
at the level of thin-homotopy classes of paths.
That is, there exists a unique group element $\beta'^{-1} * \beta - \beta'(0) \in P'_0\bbR^n$ whose action transforms $\beta'$ into $\beta$ (up to thin homotopy).
This establishes the map $\ev_1 \colon P'\bbR^n \to \bbR^n$ as a principal bundle on $\bbR^n$ with structure group $P'_0\bbR^n$.

We observe that the bundle $P'\bbR^n \to \bbR^n$ comes endowed with two canonical structures: a parallel transport and a global section.
To describe the parallel transport, suppose that we have any smooth path $\gamma$ in the base manifold $\bbR^n$, together with a point $[\beta] \in P'\bbR^n$ in the fibre over $\gamma(0)$.
That is, $\beta$ is a smooth path which terminates at the initial point of $\gamma$.
Let $\gamma_{[0,t]}$ denote the restriction of $\gamma \colon [0,1] \to \bbR^n$ to the time interval from $0$ to $t \in [0,1]$.
Using the notation of Section~\ref{sec:PT of connection}, the parallel transport along $\gamma$ starting at $p = \beta$ is now given by the thin homotopy class of the smooth path
\begin{equation}
    \label{eq:PT in P'R^n}
    \hat{\gamma} \colon [0,1] \longrightarrow P' \bbR^n\,,
    \quad
    t \longmapsto \gamma_{[0,t]} * \beta\,.
\end{equation}
This prescription satisfies all the properties we would expect of a parallel transport, including that $\hat{\gamma}(1)$ depends only on the thin homotopy class of $\gamma$.

Our task now is, therefore, to reverse-engineer from the parallel transport~\eqref{eq:PT in P'R^n} the connection that induces it, or a differential-form representative thereof.
We utilise the second natural geometric structure present on the bundle $\ev_1 \colon P'\bbR^n \to \bbR^n$: its global section.
This is the map
\begin{equation}
    s \colon \bbR^n \longrightarrow P'\bbR^n
\end{equation}
which sends a point $x \in \bbR^n$ to the constant path $s(x)$ at the point $x$, i.e.~$s(x)(t) = x$ for each $t \in [0,1]$.

By Equation~\eqref{eq:group-valued PT via section}, we can now extract a representation of the parallel transport which is valued in the structure group $P'_0\bbR^n$.
For an arbitrary path $\gamma \colon [0,1] \to \bbR^n$, we have that
\begin{align}
\label{eq: path-bundle PT wrt to s}
    \hat{\gamma}(t) &= \gamma_{[0,t]} * s(\gamma(0))
    \\
    &= \gamma_{[0,t]}
    \notag
    \\
    &= s(\gamma(t)) * \gamma_{[0,t]}
    \notag
    \\
    &= s(\gamma(t)) \triangleleft \underbrace{(\gamma_{[0,t]} - \gamma(t))}_{=g(t) \in P'_0 \bbR^n}\,,
    \notag
\end{align}
where both sides of this identity should be taken modulo thin homotopies.
In the first identity, we used the definition of parallel transport, in the second and third that there are thin homotopies between $\gamma_{[0,t]} * s(\gamma(0))$, $\gamma_{[0,t]}$ and $s(\gamma(t)) * \gamma_{[0,t]}$, and in the final identity, we rewrote the concatenation as the action of an element of the structure group.

The central observation about Equation~\eqref{eq: path-bundle PT wrt to s}---and the answer to our guiding question for a parallel transport, or connection, that can distinguish paths up to thin homotopy---is that the group element $g(t)$ which represents the parallel transport along $\gamma$ with respect to the section $s$ \textit{is the path $\gamma$ itself, up to a translation}.
We have thus derived the following:

\begin{center}
    \textit{The parallel transport described in Equations~\eqref{eq:PT in P'R^n} and~\eqref{eq: path-bundle PT wrt to s} distinguishes paths in $\bbR^n$ up to thin homotopy, reparameterization and translation}!
\end{center}

Next, we need to determine the differential form representing the connection of this parallel transport with respect to the section $s$.
Equations~\eqref{eq:PT and connection} and~\eqref{eq:g(Delta t)---first step} hold the key to achieving that.
For ease of notation, we write $\dot{\gamma}(0) = w$ for the tangent vector to $\gamma$ at time $0$.
We may think of $w$ as representing an infinitesimally short, straight path, starting at a point $x \in \bbR^n$ and running in the direction of a $v$, i.e.~a path from $x$ to $x + \Delta t \cdot v + \cO(\Delta t^2)$.
In the notation of Equation~\eqref{eq:PT and connection} and~\eqref{eq:g(Delta t)---first step}, we have that $g(0)$ is the unit element in the structure group $P_0'\bbR^n$, i.e.~the thin homotopy class of the constant path at the origin $0 \in \bbR^n$.

We remark that the total space $P'\bbR^n$ and the structure group $P'_0\bbR^n$ carry compatible vector space structures induced by point-wise addition and rescaling of paths, i.e.
\begin{equation}
    (\lambda \gamma + \mu \gamma')(t) = \lambda \cdot \gamma(t) + \mu \cdot \gamma'(t)\,.
\end{equation}
This is also compatible with taking thin-homotopy classes of paths.
Note that taking such linear combinations changes the endpoints of paths unless both paths end at the origin (in particular, it does not preserve the fibres of the map $\ev_1 \colon \bbR^n \to \bbR^n$).
It follows, though, that the vector space underlying the Lie algebra of $P'_0 \bbR^n$ can be identified again with $P'_0 \bbR^n$.

We now compute, in the spirit of Equation~\eqref{eq:g(Delta t)---first step}, the value of the parallel transport $\hat{\gamma}(\Delta t)$ at first order in $\Delta t$.
We find that
\begin{align}
    \hat{\gamma}(\Delta t)
    &= s \big( \gamma(\Delta t) \big) \cdot g(\Delta t)
    \\
    &= \mathrm{const}_{(x + \Delta t\, w)} \triangleleft \hat{w}\,,
    \notag
\end{align}
where $\hat{w}$ is the first-order path from $\Delta t \cdot w$ to the origin.
Thus, we obtain that $g(\Delta t) = \hat{w}$.
At the same time, we have that
\begin{align}
\label{eq: value of signature connection}
    A(w) &= A(w) \, c_0
    = \dot{g}(0)
    = \frac{g(\Delta t) - g(0)}{\Delta t}
    \\
    &= \frac{1}{\Delta t} (\hat{w} - c_0)
    = \hat{w}\,.
    \notag
\end{align}
Equation~\eqref{eq: value of signature connection} is at the very heart of the path signature.
It almost fully identifies the values of the 1-form which presents the parallel transport we introduced in this section with respect to the section $s$.
It says that, when evaluated on a tangent vector $w \in \bbR^n$, the 1-form returns precisely that tangent vector.
The only, rather conceptual, thing remaining is to identify the Lie algebra which we should interpret this value $A(w) = w$ to live in (since $A$ must be a Lie-algebra valued 1-form).

Before proceeding with our derivation of the path signature, let us remark that the value $A(w)$ does not depend on where we are in $\bbR^n$; that is, the 1-form $A$ is constant, or translation-invariant.
This is not a coincidence:
the bundle $P'\bbR^n \to \bbR^n$ admits a natural action of the group of translations of $\bbR^n$.
Furthermore, both the parallel transport we defined and the section $s$ we employed are compatible%
\footnote{
    In mathematical terms, the bundle has an \textit{equivariant structure} for the translation action and the global section $s$ we employ is an equivariant section.
    Moreover, the parallel transport we defined also preserves the equivariant structure, and so we will obtain a translation-invariant 1-form representative.
}
with this action of the translation group on $P'\bbR^n$.

It remains to link the structure group $P'_0\bbR^n$ and the usual algebra-valued description of the signature (see Section~\ref{sec:background}).
So far, we have derived a connection form $A$ on $\bbR^n$ which sends a tangent vector $w \in \bbR^n$ to the vector $w$ itself.
However, that merely implies that $\bbR^n$ is somehow contained in the actual Lie algebra in which $A$ takes its values.
It does not identify for us the Lie algebra itself.

Here we will provide an argument that, at least, there is an embedding of the structure group $P'_0\bbR^n$ into the completed tensor algebra $\bar{T}\bbR^n$.
If such an embedding $\iota \colon P'_0\bbR^n \hookrightarrow \bar{T}\bbR^n$ exists, then because it is injective the composition of the $P'_0\bbR^n$-valued parallel transport in the bundle $P\bbR^n \to \bbR^n$ discussed above with the embedding $\iota$ will still be injective, and thus still distinguish paths in $\bbR^n$ up to reparameterization, translation and thin homotopy.

Further, we believe that the embedding $\iota$ even induces an isomorphism on Lie algebras.
However, to our knowledge, both statements about $\iota$ are currently not established mathematical facts, but rather are currently being researched (including by one of the authors).

To see why an embedding $\iota$ might exist, first consider linear paths $\gamma[w]$ associated to vectors $w \in \bbR^n$.
Whatever the structure group of our bundle, we know that the value of the connection form on the velocity vector $\dot{\gamma}[w](t) = w$ satisfies $A(\dot{\gamma}[w]) = w$ for all parameter values $t \in [0,1]$.
Thus, by~\eqref{eq: iterated integral formula for Pexp} we can write the parallel transport (or lifted path) as
\begin{equation}
    \hat{\gamma}[w] = \exp_{P'_0\bbR^n}(w)\,,
\end{equation}
where the exponential is taken, at least formally, in the structure group $P'_0\bbR^n$.
By the functoriality of parallel transport, we also obtain an expression for the parallel transport along piecewise linear paths:
any piecewise linear path $\gamma$ in $\bbR^n$ can be described by a sequence $(w_1, \ldots, w_N)$ of vectors $w_1, \ldots, w_N \in \bbR^n$.
Up to possible reparameterization, $\gamma$ can be written as the concatenation
\begin{equation}
    \gamma = \gamma[w_N] * \cdots * \gamma[w_1]\,.
\end{equation}
Thus, its parallel transport can be written, at least formally, as
\begin{equation}
    \hat{\gamma} = \exp_{P'_0\bbR^n}(w_N) \cdots \exp_{P'_0\bbR^n}(w_1)
    \quad \in P'_0\bbR^n\,.
\end{equation}
Note that since the exponential is taken in a nonabelian group, we are not allowed to change the order of factors in this product.

To obtain the embedding $\iota \colon P'_0\bbR^n \hookrightarrow \bar{T}\bbR^n$, the idea is to send a piecewise linear path $\gamma$ as above to the product
\begin{equation}
    \exp_\otimes(w_N) \otimes \cdots \otimes \exp_\otimes(w_1)
    \quad \in \bar{T}\bbR^n
\end{equation}
of the corresponding exponentials in the completed tensor algebra $\bar{T}\bbR^n$.

We have to argue that this prescription produces an injective map from piecewise linear paths to the (invertible elements in) the completed tensor algebra $\bar{T}\bbR^n$.
First, the only linear path $\gamma[w]$ that will be sent to the unit $1 \in \bar{T}\bbR^n$ is the constant linear path $\gamma[0]$.
Next, consider a concatenation of two linear paths, i.e.~$\gamma = \gamma[w_2] * \gamma[w_1]$.
This will be sent to the product $\exp_\otimes(w_2) \otimes \exp_\otimes(w_1)$ in $\bar{T}\bbR^n$.
We can expand this sum as
\begin{equation}
    \exp_\otimes(w_2) \otimes \exp_\otimes(w_1)
    = \sum_{k = 0}^\infty \sum_{l = 0}^k \binom{l}{k} w_2^{\otimes k-l} \otimes w_1^{\otimes l}\,.
\end{equation}
This turns out to agree with
\begin{equation}
    \exp_\otimes(w_2 + w_1) = \sum_{k = 0}^\infty \frac{1}{k!} (w_2 + w_1)^{\otimes k}
\end{equation}
precisely if $w_2$ and $w_1$ are collinear, i.e.~are scalar multiples of each other, since it is precisely in that case that $w_2 \otimes w_1 = w_1 \otimes w_2$.
In that case, consequently, the images in $\bar{T}\bbR^n$ of the paths $\gamma[w_2] * \gamma[w_1]$ and $\gamma[w_2+w_1]$ agree, potentially breaking the injectivity of the map $P'_0\bbR^n \to \bar{T}\bbR^n$.
However, it is also true that the paths $\gamma[w_2] * \gamma[w_1]$ and $\gamma[w_2+w_1]$ are thin homotopy invariant if and only if $w_2$ and $w_1$ are collinear.

The subgroup $PL'_0 \bbR^n \subset P'_0\bbR^n$ of the piecewise linear paths in $\bbR^n$ ending at the origin, modulo thin homotopy, can equivalently be described as follows.
Above, we already realized that any piecewise linear path $\gamma$ can be encoded by a finite sequence $(w_1, \ldots, w_N)$ of vectors in $\bbR^n$.
Concatenation of such paths corresponds to concatenating such sequences of vectors.
The thin homotopy relation on paths is reflected by the relation that
\begin{align}
    &(w_N, \ldots, w_{l+2}, w_{l+1}, w_l, w_{l-1}, \ldots, w_1)
    \\
    &= (w_N, \ldots, w_{l+2}, w_{l+1} + w_l, w_{l-1}, \ldots, w_1)
    \notag
\end{align}
whenever the vectors $w_{l+1}$ and $w_l$ are collinear.
Further, the relation that $\gamma[0]$ is the neutral element for path concatenation corresponds to the relation that any $0$-vector can be cancelled from any sequence of vectors.

By the arguments about sending piecewise linear paths in $\bbR^n$ to products of exponentials in $\bar{T}\bbR^n$, we now observe that the induced map
\begin{equation}
    (w_N, \ldots, w_1) \longmapsto \exp_\otimes(w_N) \otimes \cdots \otimes \exp_\otimes(w_1)
\end{equation}
respects precisely those two relations on sequences of vectors (or, equivalently, the piecewise linear paths the sequences represent).
Thus, we obtain an embedding $PL'_0 \bbR^n \hookrightarrow \bar{T}\bbR^n$ which is compatible with the multiplication and sends the unit element to the unit element.

Finally, we extend this map to all elements of $P'_0\bbR^n$ by using that any smooth path in $\bbR^n$ can be approximated by piecewise linear ones.
Thus, we obtain that the composition of the canonical parallel transport on the bundle $P'\bbR^n \to \bbR^n$ discussed in this section with the group map $P'_0\bbR^n \hookrightarrow \bar{T}\bbR^n$ is a map which can distinguish paths in $\bbR^n$ up to translation, reparameterization and thin homotopy.
We also see from~\eqref{eq: signature of linear path} that this map agrees with the signature on linear paths.
The functoriality of the signature then implies that the map also agrees with the signature on piecewise linear paths, and continuity of both parallel transport and the signature then imply that the map we have described in this appendix is, indeed, the same as the path signature on all piecewise smooth paths.

\bsp	
\label{lastpage}
\end{document}